\documentclass[a4paper,oneside,10pt]{amsart}

\usepackage[T1]{fontenc}
\usepackage{amsmath,amssymb,epsf}
\usepackage{amsfonts,amsbsy,amscd,stmaryrd}
\usepackage{latexsym,amsopn}
\usepackage{amsthm}
\usepackage{esint}
\usepackage{mathrsfs}

\usepackage{graphicx}
\usepackage{color}
\definecolor{Blue}{rgb}{0.,0.,1.}
\definecolor{Red}{rgb}{1.,0.,0.}

\newcounter{smallarabics}
\newenvironment{arabicenumerate}
{\begin{list}{{\normalfont\textrm{(\arabic{smallarabics})}}}
  {\usecounter{smallarabics}\setlength{\itemindent}{0cm}
   \setlength{\leftmargin}{5ex}\setlength{\labelwidth}{4ex}
   \setlength{\topsep}{0.75\parsep}\setlength{\partopsep}{0ex}
   \setlength{\itemsep}{0ex}}}
{\end{list}}

\newcounter{smallroman}

\let\origmaketitle\maketitle
\def\maketitle{
  \begingroup
  \def\uppercasenonmath##1{} 
  \let\MakeUppercase\relax 
	\origmaketitle
  \endgroup
}

\newcommand{\ben}{\begin{arabicenumerate}}  
\newcommand{\een}{\end{arabicenumerate}}

\def\init{\setcounter{equation}{0}}

\newtheorem{theorem}{Theorem}[section]

\newtheorem{proposition}[theorem]{Proposition}
\newtheorem{lemma}[theorem]{Lemma}
\newtheorem{definition}[theorem]{Definition}

\newtheorem{remark}[theorem]{Remark}
\newtheorem{example}[theorem]{Example}
\newcommand{\beq}{\begin{equation}}
\newcommand{\eeq}{\end{equation}}

\newcommand{\bex}{\begin{example}}
\newcommand{\eex}{\end{example}}
\def\bel{\begin{lemma}}
\def\eel{\end{lemma}}
\def\bet{\begin{theorem}}
\def\eet{\end{theorem}}
\def\bed{\begin{definition}}
\def\eed{\end{definition}}
\def\ber{\begin{remark}}
\def\eer{\end{remark}}

\def\rr{{\mathbb R}}

\def\cc{{\mathbb C}}
\def\nn{{\mathbb N}}

\def\slim{{\rm s-}\lim}
\def\wlim{{\rm w-}\lim}

\def\bar{\overline}

\def\cinf{C^\infty}
\def\c0inf{C_0^\infty}

\def\proof{
\noindent{\bf Proof.}\ \ }

\usepackage{calrsfs}
\DeclareMathAlphabet{\pazocal}{OMS}{zplm}{m}{n}
\DeclareMathAlphabet{\mathsfsl}{OMS}{cmss}{m}{n}

\DeclareSymbolFont{altletters}  {OML}{zplm}{m}{it}
\DeclareMathSymbol{\altdelta}{\mathalpha}{altletters}{"0E}
\DeclareMathSymbol{\alteta}{\mathalpha}{altletters}{"11}

\def\cY{{\pazocal Y}}

\def\cS{{\pazocal S}}

\def\cD{{\mathcal D}}

\def\cC{{\pazocal C}}

\def\CCR{{\rm CCR}}

\def\i{{\rm i}}
\def\ad{{\rm ad}}

\DeclareMathOperator{\Dom}{Dom}

\def\qed{$\Box$\medskip}

\DeclareMathOperator{\Ker}{Ker}

\def \p{ \partial}
\def\12{\frac{1}{2}}
\def\14{\frac{1}{4}}

\def\supp{{\rm supp}}
\def\e{{\rm e}}

\def\d{{\rm d}}
\newcommand{\one}{\boldsymbol{1}}
\def\cH{{\pazocal H}}

\def\coinf{C_{\rm c}^\infty}

\def\c{{\pazocal }}

\def\cX{{\pazocal X}}

\def\12{\frac{1}{2}}

\def\supp{{\rm supp}}

\def\ad{{\rm ad}}
\def\e{{\rm e}}

\def\bep{\begin{proposition}}
\def\eep{\end{proposition}}

\newcommand{\mat}[4]{\begin{pmatrix}#1 &#2  \\ #3 &#4 \end{pmatrix}}
\newcommand{\col}[2]{\begin{pmatrix}#1 \\#2\end{pmatrix}}

\def\CARal{{\rm C\hskip 0.25 em \hbox{\raise 1.72 ex 
\hbox{$\scriptscriptstyle\rm al$}\kern -0.57 em A}R}}

\def\otimesal{\mathop{\hbox{\raise 1.5 ex
  \hbox{$\scriptscriptstyle\rm al$}
\kern -0.92 em \hbox{$\otimes$}}}}
\def\oplusal{\mathop{\hbox{\raise 1.5 ex
  \hbox{$\scriptscriptstyle\rm al$}
\kern -0.92 em \hbox{$\oplus$}}}}
\def\Gammal{\hbox{\raise 1.68 ex 
\hbox{$\scriptscriptstyle\rm al$}\kern -0.50 em $\Gamma$}}
\def\Bal{\hbox{\raise 1.68 ex 
\hbox{$\scriptscriptstyle\rm  al$}\kern -0.50 em $B$}}
\def\CARal{{\rm C\hskip 0.25 em \hbox{\raise 1.72 ex 
\hbox{$\scriptscriptstyle\rm al$}\kern -0.57 em A}R}}

\def\cE{\pazocal{E}}

\DeclareMathAlphabet{\mathpzc}{OT1}{pzc}{m}{it}

\DeclareSymbolFont{boldoperators}{OT1}{cmr}{bx}{n}
\SetSymbolFont{boldoperators}{bold}{OT1}{cmr}{bx}{n}

\makeatletter
\newcommand*{\defeq}{\mathrel{\rlap{%
                     \raisebox{0.3ex}{$\m@th\cdot$}}%
                     \raisebox{-0.3ex}{$\m@th\cdot$}}%
                     =}
                     
\makeatother
\makeatletter
\newcommand*{\eqdef}{=\mathrel{\rlap{%
                     \raisebox{0.3ex}{$\m@th\cdot$}}%
                     \raisebox{-0.3ex}{$\m@th\cdot$}}%
                     }
\makeatother

\def\Texp{{\rm Texp}}
\def\cA{{\mathcal A}}

\DeclareMathAlphabet{\mathpzc}{OT1}{pzc}{m}{it}

\newcommand{\bea}{\begin{aligned}}
\newcommand{\beal}{\begin{array}{l}}
\newcommand{\eeal}{\end{array}}
\newcommand{\eea}{\end{aligned}}

\newcommand{\traa}[1]{\mskip-6mu\upharpoonright_{#1}}
\def\pe{\overline{\p}}

\def\diag{{\rm diag}}

\def\BT{{\rm BT}}

\def\dito{\!\cdot\!}
\newcommand{\tra}[1]{\mskip-6mu\upharpoonright_{#1}\mskip+4mu}
\makeatletter
\begin{document}
\title[On the adiabatic limit of Hadamard states]{\Large On the adiabatic limit of Hadamard states}
\author{}

\author{\normalsize Nicol\`o \textsc{Drago} \& Christian \textsc{G\'erard} }
\address{Dipartimento di Matematica, Universit\`a di Genova - Via Dodecaneso 35, I-16146 Genova, Italy}
\email{drago@dima.unige.it}
\address{Universit\'e Paris-Sud XI, D\'epartement de Math\'ematiques, 91405 Orsay Cedex, France}
\email{christian.gerard@math.u-psud.fr}
\thanks{The authors would like to thank the referees for their help in improving the readability of the paper}
\begin{abstract}
We consider the adiabatic limit of Hadamard states for free quantum Klein-Gordon fields, when the background metric and  the field mass  are slowly varied from their initial to final values.  If the Klein-Gordon field stays massive, we prove that the adiabatic limit of the  initial vacuum state is the  (final) vacuum state, by extending to the symplectic framework the adiabatic theorem of Avron -Seiler-Yaffe. 

In cases when only the field mass is varied,  using an abstract version of the mode decomposition method we can also consider the case when the initial or final mass vanishes, and the initial state is either a thermal state or a more general Hadamard state.  
\end{abstract}

\maketitle
\section{Introduction}\label{intro}\init
In this paper we study the adiabatic limit of Hadamard states for free quantum Klein-Gordon fields. Hadamard states play nowadays a crucial role in the algebraic approach to Quantum Field Theory on curved spacetimes.  They are suitable linear positive and normalized functionals on the $*$-algebra of observables \cite{KM14}, which enjoy further microlocal properties \cite{Ra96a,Ra96b}. They play an important role in algebraic Quantum Field Theory for several reasons, \cite{GK89, Wa94,FV13,HW02}, ultimately linked to the  fact  that the Hadamard condition is the correct criterion to single out physically relevant states.
Nowadays the literature on Hadamard states is wide, ranging from existence results \cite{FNW81,FSW78} to explicit constructing techniques \cite{BDM14,DD15,DMP06,FMR15,GW14a,GW14b,WZ14}.

In this paper we describe another construction of Hadamard states via a deformation procedure in parameter space.
This deformation procedure is obtained by considering an ``intermediate'' theory with a smoothly deformed parameter, which interpolates between the two values of interest (eg between two values of the mass).
States for this latter theory can be thought as smooth deformations of states from one theory to the other.
Actually, it is only in the final step that one really recovers a state for the theory of interest: This step consists in a limit procedure, the so-called {\em adiabatic limit}.

In the first part of the paper we consider  massive Klein-Gordon  field with an external electromagnetic potential in a globally hyperbolic spacetime. The metric, electromagnetic potential and the field mass are smoothly deformed from their initial to final values.
We show in Thm. \ref{adiabatic-limit} that the adiabatic limit of  the    initial vacuum state is  again the final vacuum state.   For this we generalize the well known results of \cite{ASY} on the adiabatic limit  for a symplectic, rather than unitary, dynamics.

The previous analysis leaves out the massless case, which is typically affected by infrared divergences.
The treatment of this case is the content of the second part of the paper, which specializes the model previously described to the case where only the field mass is varied.   The Klein-Gordon equation is then separable  and one can restrict attention to quasi-free states whose covariances are diagonal w.r.t. the spatial Laplacian. The construction  of such states is known in the physics literature as the mode decomposition method. Some aspects of this analysis already appear, in a different formulation and in special cases, in \cite{DD15,DHP15}.

The main result proved in Prop.  \ref{p3.1} is that the adiabatic limit for the mass parameter can be performed for a large class of  such states,   containing in particular vacuum states  and thermal states.
As a particular case we prove that the KMS property, which characterizes states in thermal equilibrium (see \cite{Sa13}), is not preserved by the adiabatic limit. 

The paper is structured as follows: Sections \ref{sec3}-\ref{Section: Adiabatic limits of quasi-free states} are devoted to recollect some well-known material and to formulate precisely the problem of the adiabatic limit for the model of interest.  Section \ref{sec1} deals with the adiabatic Theorem for symplectic dynamics (Thm.\ref{theoadia}) which generalizes the result of \cite{ASY} to the symplectic case.  Finally Section \ref{sec4} deals with the massless to massive transition, analyzing in particular the Hadamard property of the adiabatic limit as well as the adiabatic limit of vacuum and KMS states.

\subsection{Notation}\label{sec0.not}

- we set $\langle x\rangle= (1+ x^{2})^{\12}$ for $x\in \rr^{n}$.

- the domain of a closed, densely defined operator  $a$  will be denoted by $\Dom a$ and equipped with the graph norm, its resolvent set by $\rho(a)$.

- if $a$ is a selfadjoint operator on a Hilbert space $\cH$, we write $a>0$ if $a\geq 0$ and $\Ker a=\{0\}$.  We set $\cS=\{u\in \cH : u= \one_{[\delta, R]}(a)u, \ \delta, R>0\}$. For $s\in\rr$ we denote by $\langle a\rangle^{s}\cH$ the completion of $\cS$ for the norm $\| u\|_{-s}=\| \langle a\rangle ^{-s}u\|$.  Similarly  if $a>0$ we denote by $a^{s}\cH$ the completion of $\cS$ for the norm $\|u\|= \|a^{-s}u\|$.  

- functions of $a$ will be denoted by $f(a)$, in particular if $\Delta\subset \rr$ is a Borel set, $\one_{\Delta}(a)$ denotes the spectral projection on $\Delta$ for $a$.

- if $\rr\ni t\mapsto b(t)$ is a map with values in closed densely defined operators on $\cH$, satisfying the conditions of Kato's theorem,   see \cite[Thm. X.70]{RS2} or \cite{SG} for a recent exposition,  the strongly continuous two parameter group with generator $b(t)$ will be denoted by  $\Texp(\i \int_{s}^{t}b(\sigma)d\sigma)$.

- the operator of multiplication by a function $f$ will be denoted by $f$, while the operators of partial differentiation will be denoted by $\pe_{i}$, so that $[\pe_{i}, f]= \p_{i}f$.

\section{Free quantized Klein-Gordon fields}\label{sec3}\init
We now briefly recall some background material on free quantized Klein-Gordon fields, referring  for example to \cite{BGP, KM14} for details.  We adopt the framework of {\em charged fields}, corresponding to complex solutions of  the Klein-Gordon equation, which we find more convenient. We refer the reader to \cite[Sect. 1]{GW14a} for details.
\subsection{Charged bosonic fields}
 In this framework the phase space, used to construct the CCR algebra, is a pseudo-unitary space $(\cY, q)$, ie $\cY$ is a complex vector space and $q\in L_{\rm h}(\cY, \cY^{*})$ a non-degenerate hermitian form, instead of a real symplectic space $(\cX, \sigma)$ as usual. 
Let  hence $\cY$ a complex vector space, $\cY^{*}$ its anti-dual. Sesquilinear forms on $\cY$ are identified with elements of $L(\cY, \cY^{*})$ and the action of a sesquilinear form $\beta$ is correspondingly denoted by $\bar{y}_{1}\dito \beta y_{2}$ for $y_{1}, y_{2}\in \cY$. We fix  $q\in L_{\rm h}(\cY, \cY^{*})$  a non degenerate hermitian form on $\cY$.

  The   $^{*}-$algebra $\CCR(\cY, q)$ is the (complex) $^{*}-$algebra generated by symbols $\one, \psi(y), \psi^{*}(y),y\in \cY$ and the relations:
\[
\begin{array}{l}
\psi(y_{1}+ \lambda y_{2})= \psi(y_{1})+ \bar{\lambda}\psi(y_{2}), \ y_{1}, y_{2}\in \cY, \lambda\in \cc,\\[2mm]
\psi^{*}(y_{1}+ \lambda y_{2})= \psi^{*}(y_{1})+ \lambda\psi^{*}(y_{2}), \ y_{1}, y_{2}\in \cY, \lambda\in \cc,\\[2mm]
[\psi(y_{1}), \psi(y_{2}]= [\psi^{*}(y_{1}), \psi^{*}(y_{2})]=0, \ [\psi(y_{1}), \psi^{*}(y_{2})]= \bar{y}_{1}\dito qy_{2}\one, \ y_{1}, y_{2}\in\cY,\\[2mm]
\psi(y)^{*}= \psi^{*}(y), \ y\in \cY.
\end{array}
\]
 A state $\omega$ on $\CCR(\cY, q)$ is {\em (gauge invariant) quasi-free} if 
 \[
\omega(\prod_{i=1}^{p}\psi(y_{i})\prod_{i=1}^{q}\psi^{*}(y_{j}))=\left\{  \begin{array}{l}
0\hbox{ if }p\neq q,\\
\sum_{\sigma\in S_{p}}\prod_{i=1}^{p}\omega(\psi(y_{i})\psi^{*}(y_{\sigma(i)}))\hbox{ if }p=q.
\end{array} \right.
\]
There is no loss of generality to restrict oneself to charged fields and gauge invariant states, see eg the discussion in \cite[Sect. 2]{GW14a}. 
It is convenient to associate to  $\omega$ its {\em (complex) covariances} $\lambda_{\pm}\in L_{\rm h}(\cY, \cY^{*})$ defined by:
\[
\begin{array}{l}
\omega(\psi(y_{1})\psi^{*}(y_{2}))\eqdef \bar{y}_{1}\dito \lambda_{+} y_{2}, \\[2mm]
 \omega(\psi^{*}(y_{2})\psi(y_{1}))\eqdef \bar{y}_{1}\dito \lambda_{-} y_{2},
 \end{array} \ y_{1}, y_{2}\in \cY.
\]
It is well-known that  two  hermitian  forms $\lambda_{\pm}\in L_{\rm h}(\cY, \cY^{*})$ are the covariances of a quasi-free state $\omega$ iff
 \begin{equation}
\label{ef.1}
\lambda_{\pm}\geq 0, \ \lambda_{+}- \lambda_{-}=q.
\end{equation}

\subsection{Free quantized Klein-Gordon fields}
Let  $(M, g)$ be a globally hyperbolic spacetime, $A_{a}(x)dx^{a}$ a smooth $1-$form on $M$ and $m\in \coinf(M; \rr)$ a smooth real function.  We  set
\beq\label{defdekg}
P=-(\nabla^{a}-\i A^{a}(x))(\nabla_{a}- \i A_{a}(x))+ m(x)
\eeq
the associated Klein-Gordon operator. Let  $G^{\pm}$  be the advanced/retarded inverses of $P$ and  $G\defeq G^{+}- G^{-}$ the causal propagator.  Denote by ${\rm Sol}_{\rm sc}(KG)$ the space of smooth,  complex, space-compact solutions of the Klein-Gordon equation $P\phi=0$.

We equip ${\rm Sol}_{\rm sc}(KG)$ with the hermitian form
\[
\overline{\phi}\dito q \phi\defeq\i\int_{\Sigma}\left(\overline{(\p_{a}-\i A_{a})\phi} \phi- \overline{\phi}(\p_{a}- \i A_{a})\phi\right)n^{a}ds_{\Sigma}
\]
where $\Sigma$ is a spacelike Cauchy hypersurface, $n^{a}$ is the future directed normal to $\Sigma$ and $ds_{\Sigma}$ the induced density on $\Sigma$.  The above expression  is independent on the choice of $\Sigma$ and $({\rm Sol}_{\rm sc}(KG),q)$ is a pseudo-unitary  space, i.e. $q$ is non degenerate.
 
It is well-known that the sequence
\[
0\longrightarrow \coinf(M)\mathop{\longrightarrow}^{P} \coinf(M)\mathop{\longrightarrow}^{G} {\rm Sol}_{\rm sc}(KG)\mathop{\longrightarrow}^{P}0
\]
is exact and
\[
\overline{Gu}\dito q  Gu= \i^{-1}(u| G u)_{M}\eqdef \overline{[u]}\dito Q [u], \ [u]\in\frac{\coinf(M)}{P \coinf(M)},
\]
where $(u|v)_{M}= \int_{M}\bar{u}vdVol_{g}$. It follows that
\[
( \frac{\coinf(M)}{P\coinf(M)}, Q)\mathop{\longrightarrow}^{G}({\rm Sol}_{\rm sc}(KG), q)
\]
is an isomorphism of pseudo-unitary spaces.  Fixing a space-like Cauchy hypersurface $\Sigma$ and setting
\[
\rho: \cinf_{\rm sc}(M)\ni \phi\mapsto \rho\phi= \col{\phi\tra\Sigma}{n^{a}(\i^{-1}\p_{a}\phi-A_{a}\phi)\tra\Sigma}= f\in\coinf(\Sigma)\otimes \cc^{2},
\]
we obtain, since the  Cauchy problem 
\[
\left\{
\begin{array}{l}
P\phi=0, \\
\rho u=f\end{array}
\right.
\]
for $f\in\coinf(\Sigma)\otimes \cc^{2}$  that
\[
({\rm Sol}_{\rm sc}(KG), q)\mathop{\longrightarrow}^{\rho}(\coinf(\Sigma)\oplus\coinf(\Sigma), q)
\]
is pseudo-unitary, where
\beq\label{calim}
\bar{f}\dito qf=\int_{\Sigma}\bar{f}_{1}f_{0}+ \bar{f}_{0}f_{1}ds_{\Sigma}, \ f= \col{f_{0}}{f_{1}}.
\eeq
\subsection{Quasi-free states}
One restricts attention to quasi-free states on $\CCR(\cY, q)$ whose covariances are given by distributions on $M\times M$, ie such that there exists $\Lambda^{\pm}\in \cD'(M\times M)$ with
\beq\label{ef.2b}
\begin{array}{l}
\omega(\psi([u_{1}])\psi^{*}([u_{2}]))= (u_{1}| \Lambda^{+}u_{2})_{M}, \\[2mm]
 \omega(\psi^{*}([u_{2}])\psi([u_{1}]))= (u_{1}| \Lambda^{-}u_{2})_{M},
\end{array}
 \ u_{1}, u_{2}\in \coinf(M).
\eeq
In the sequel the distributions $\Lambda^{\pm}\in \cD'(M\times M)$ will be called the {\em spacetime covariances} of  the state $\omega$.

In \eqref{ef.2b} we identify distributions on $M$ with distributional densities using the density $dVol_{g}$ and use hence the notation $(u|\varphi)_{M}$, $u\in \coinf(M)$, $\varphi\in \cD'(M)$ for the duality bracket. We have then
\[
\begin{array}{l}
P(x, \p_{x})\Lambda^{\pm}(x, x')= P(x', \p_{x'})\Lambda^{\pm}(x, x')=0,\\[2mm]
 \Lambda^{+}(x, x')- \Lambda^{-}(x, x')= \i^{-1}G(x, x').
\end{array}
\]

Since
\[
( \frac{\coinf(M)}{P\coinf(M)}, Q)\ \mathop{\longrightarrow}^{\rho\circ G}\ (\coinf(\Sigma)\otimes \cc^{2}, q)
\]
is an isomorphism of pseudo-unitary spaces,  it  follows that   a quasi-free state with space-time covariances $\Lambda^{\pm}$ is uniquely defined by its  {\em Cauchy surface covariances} $\lambda_{\Sigma}^{\pm}$ defined by:
 \begin{equation}
\label{ef.4}
\Lambda^{\pm}\eqdef (\rho E)^{*} \lambda_{\Sigma}^{\pm}( \rho E).
\end{equation}
 Using the canonical scalar product  $(f|f)_{\Sigma}\defeq\int_{\Sigma} \bar{f}_{1}f_{1}+ \bar{f}_{0}f_{0}d\sigma_{\Sigma}$ we identify $\lambda_{\Sigma}^{\pm}$ with operators, still denoted by $\lambda_{\Sigma}^{\pm}$, belonging to $L(\coinf(\Sigma)\otimes \cc^{2},\cD'(\Sigma)\otimes \cc^{2})$.
 
A pair $\lambda_{\Sigma}^{\pm}$ of hermitian forms on $\coinf(\Sigma)\otimes \cc^{2}$  is the pair of Cauchy surface covariances of a quasi-free state iff
\begin{equation}
\label{calum}
\lambda_{\Sigma}^{\pm}\geq 0, \ \lambda_{\Sigma}^{+}- \lambda_{\Sigma}^{-}=q,
\end{equation}
where the charge $q$ is defined in \eqref{calim}.

\section{Adiabatic limits of quasi-free states}\label{Section: Adiabatic limits of quasi-free states}\init
In this section we formulate the problem that we will consider in this paper, namely the existence of adiabatic limits
for quasi-free states. The formulation relies on a $1+d$ decomposition, ie on fixing some time coordinate.

We also state Thm. \ref{adiabatic-limit} about the adiabatic limits of {\em vacuum states}.
\subsection{$1+d$ decompositions}\label{sec1.tot}
 We  consider  simple model spacetimes $M= \rr\times \Sigma$ equipped with the Lorentzian metric
 \[
g= - dt^{2}+ h_{ij}(t, x)dx^{i}dx^{j},
\]  
where $\Sigma$ is a smooth manifold and $h_{ij}(t, x)dx^{i}dx^{j}$ is a smooth, time-dependent family of complete Riemannian metrics on $\Sigma$. We also fix a smooth $1-$form $A= V(t, x)dt+ {\rm A}_{i}(t, x)dx^{i}$ and a real function $m\in \coinf(M)$.
 We denote by $\tilde{P}$ the associated Klein-Gordon operator  as \eqref{defdekg}:
 
 \beq\label{e1.001}
\tilde{P}=(\pe_{t}- \i V(t))^{2}+ r(t)(\pe_{t}- \i V(t))+ a(t,x, \pe_{x}),
\eeq
where $V(t),r(t)$ are the operators of multiplication by $V(t, x), |h_{t}|^{-\12}\p_{t}|h_{t}|^{\12}(x)$ and
\[
\begin{array}{rl}
&\tilde{a}(t, x,\pe_{x})\\[2mm]
=& - |h_{t}|^{-\12}(x)(\pe_{j}- \i {\rm A}_{j}(t, x))|h_{t}|^{\12}(x)h_{t}(x)^{jk}(\pe_{k}- \i {\rm A}_{k}(t,x))+ m(t,x),
\end{array}
\]
 is formally selfadjoint on  $\cH_{t}= L^{2}(\Sigma, |h_{t}| dx)$.

 It is convenient to equip $\Sigma$ with the time-independent density $|h_{0}|^{\12}dx$ and to set
 \[
c^{2}(t,x):= |h_{t}|^{-\12}|h_{0}|^{\12}(x).
\]
Using the unitary transformation
\[
U: L^{2}(M, | h_{0}|^{\12}dxdt)\ni \phi\mapsto \psi= c\phi\in L^{2}(M, |h_{t}|^{\12}dxdt)
\]
for $c^{2}=|h_{t}|^{-\12}|h_{0}|^{\12}$, 
we transform $\tilde{P}$ into $P= U \tilde{P}U^{-1}= c^{-1}\tilde{P}c$. 

Using that $(\pe_{t}-\i V)c= c(\pe_{t}-\i V)+ \p_{t}c$ and $r= - 2c^{-1}\p_{t}c$, we obtain after an easy computation  that:
\begin{equation}
\label{e1.001}
P= (\pe_{t}- \i V)^{2}+a(t, x, \pe_{x}), 
\end{equation}
\[
a(t, x, \pe_{x})=a(t)= c^{-1}\tilde{a}(t, x, \pe_{x})c+ c^{-1}\p_{t}^{2}c- 2 (c^{-1}\p_{t}c)^{2},
\]
which is formally selfadjoint on $\cH= L^{2}(\Sigma, |h_{0}|^{\12}dx)$. The conserved charge for the solutions of $P \phi=0$ is:
\[
 \bar{\phi}q\phi\defeq\int_{\Sigma}\left((\overline{\i^{-1}\p_{t}\phi(t)- V(t)\phi(t)} )\phi(t)+ \overline{\phi(t)(}\i^{-1}\p_{t}\phi(t)- V(t)\phi(t))\right)|h_{0}|^{\12}dx.
\]
The corresponding identities  for causal propagators  and spacetime two-point functions  of a quasi-free state are:
 \[
G= c^{-1}\tilde{G}c, \ \Lambda^{\pm}= c^{-1} \tilde{\Lambda}^{\pm}c,
\]
and in the sequel we will consider quantized Klein-Gordon fields for $P$, instead of the original operator $\tilde{P}$, since both are equivalent.
\subsection{Assumptions}\label{sec1.assumpt}
 We will assume that  for any  interval $I\Subset \rr$ there exist constants $C_{I, n}>0$, $n\in \nn$ such that for $t\in I, x\in \Sigma$:
 \[
 \begin{array}{rl}
(H i)& C_{I, 0}^{-1}h_{0}(x)\leq h_{t}(x)\leq C_{I, 0}h_{0}(x), \\[2mm]
 (Hii)&|\p_{t}^{n}h_{t}(x)|\leq C_{I,n}h_{0}(x), \\[2mm]
 (Hiii)& |\p^{n}_{t}V(t, x)|+  |\p_{t}^{n}{\rm A}_{i}(t, x)h_{0}^{ij}(x)  \p_{t}^{n}{\rm A}_{j}(t, x)|+ |\p_{t}^{n}m(t,x)|\leq C_{I, n},\\[2mm]
 (Hiv)&\p_{i}{\rm A}_{i}(t, x)h_{0}^{ij}(x)\p_{j}{\rm A}_{j}(t,x)\leq C_{I, 0}.
  \end{array}
  \]
Let us set 
\[
a_{0}= a_{0}(x, \pe_{x})= - |h_{0}|^{-\12}\pe_{j}|h_{0}|^{\12}h_{0}^{jk}\pe_{k},
\]
which by Chernoff's theorem \cite{C} (recall that $h_{t}(x)dx^{2}$ is assumed to be  complete), is  essentially selfadjoint on $\coinf(\Sigma)$. We set $H^{1}(\Sigma):= \Dom a_{0}^{\12}$ and  $H^{-1}(\Sigma):= H^{1}(\Sigma)^{*}$ its anti-dual. We  have continuous and dense embeddings $H^{1}(\Sigma)\hookrightarrow L^{2}(\Sigma)\hookrightarrow H^{-1}(\Sigma)$.

 Similarly $a(t)$ is essentially selfadjoint on $\coinf(\Sigma)$ and    using $(H)$ we easily see that $H^{1}(\Sigma)= \Dom |a(t)|^{\12}$.

We  will need later stronger conditions than $(H)$. Setting $f^{(k)}= \p_{t}^{k}f$, we require that for $k=1,2$:
\[
\begin{array}{rl}
(Di)& a^{(k)}(t)a^{-1}(t) \hbox{  is bounded on }\cH\hbox{ locally  uniformly in  }t,\\[2mm]
(Dii)& a(t)^{\12} a^{(k)}(t)a(t)^{-3/2} \hbox{ is bounded on }\cH\hbox{ locally  uniformly in  }t,\\[2mm]
(Diii)& a(t)^{-\12}[a^{(k)}(t), a(t)]a(t)^{-1}\hbox{ is bounded on }\cH\hbox{ locally  uniformly in  }t.
\end{array}
\]
These conditions will be used in Lemma \ref{2.1} to estimate time derivatives of $a(t)^{\12}$. They are a substitute for the lack of knowledge of $\Dom a(t)$ in our abstract setting.
\begin{remark}
\def\BT{{\rm BT}}
A convenient setup where  conditions $(D)$ are satisfied is the following:  we assume that 
 $(\Sigma, h_{0})$ is of bounded geometry, see   \cite{CG85, Ro88} or \cite{GOW16} for a self contained exposition. One can  then define the spaces $\BT^{p}_{q}(\Sigma, h_{0})$ of smooth bounded $(q, p)$ tensors. If we assume  that $h\in \cinf(\rr; \BT^{0}_{2}(\Sigma, h_{0}))$, $h^{-1}\in \cinf(\rr; \BT^{2}_{0}(\Sigma,h_{0}))$, $V,m\in \cinf(\rr; \BT^{0}_{0}(\Sigma, h_{0}))$, ${\rm A}\in \cinf(\rr, \BT^{0}_{1}(\Sigma, h_{0}))$, then these assumptions are satisfied. We refer the interested reader to \cite[Sects. 2, 5]{GOW16} for  details. 
\end{remark}

\subsection{Cauchy evolution}
Denoting by
\[
\rho_{s}\phi(x)= \col{\phi(s,x)}{\i^{-1}\p_{t}\phi(s,x)- V(s, x)\phi(s, x)},\ s\in I, 
\]
the trace operator on $\Sigma_{s}= \{s\}\times \Sigma$, we know that the Cauchy problem
\beq\label{e1.1}
\left\{
\begin{array}{l}
P\phi=0,\\
\rho_{s}\phi= f\in\coinf(\Sigma)\otimes \cc^{2}
\end{array}
\right.
\eeq
is globally well posed. If $\phi= U_{s}f$ for $f\in\coinf(\Sigma)\otimes \cc^{2}$ is the solution of \eqref{e1.1},    we denote by  $U(t,s)f=\rho_{t}U_{s}f$ the Cauchy evolution for \eqref{e1.1}. 

If $\omega$ is a quasi-free state for $P$, the Cauchy surface covariances of $\omega$ for the Cauchy surface $\Sigma_{s}$ will be denoted by $\lambda^{\pm}_{s}$. Clearly we have
\[
\lambda_{t}^{\pm}= U(s,t)^{*}\lambda^{\pm}_{s}U(s,t).
\]
\subsection{Energy spaces}
Let $I\Subset \rr$ a compact interval. Let us introduce the following positivity condition:
 \[
(P)\qquad a(t, x, \pe_{x})- V^{2}(t, x)\geq C_{I}\one\hbox{ on }\cH \hbox{ for }t\in I. 
\]
In practice  $(P)$ is satisfied if we choose $m(t, x)= m^{2}$ large enough.   If $(P)$ holds 
we introduce the energy norm:
\beq\label{e1.01}
E_{t}(f, f):= (f_{1}+ V(t)f_{0}|f_{1}+ V(t)f_{0})+ (f_{0}| p(t)f_{0}),
\eeq
where $p(t)= a(t)- V^{2}(t)$ and $(u| v)$ denotes the scalar product in $\cH= L^{2}(\Sigma, |h_{0}|^{\12}dx)$.  By $(P)$ $E_{t}(\cdot, \cdot)$ is positive definite and using $(H)$ and $(P)$ we see that the norm $E_{t}(f, f)^{\12}$ is equivalent to $\| f_{0}\|_{H^{1}(\Sigma)}+ \| f_{1}\|_{L^{2}(\Sigma)}$, uniformly for $t\in I$. 

\begin{definition}\label{defoen}
 The space $H^{1}(\Sigma)\oplus L^{2}(\Sigma)$ with norm $\|f_{0}\|_{H^{1}(\Sigma)}\oplus \|f_{1}\|_{L^{2}(\Sigma)}$, resp. $E_{t}(f, f)^{\12}$ will be denoted by $\cE$, resp. $\cE_{t}$.
 \end{definition}
The norms $\|f\|_{\cE}$ and $\|f\|_{\cE_{t}}$ are uniformly equivalent for $t\in I$, using $(H)$, and $\coinf(\Sigma)\otimes \cc^{2}$ is dense in $\cE= \cE_{t}$.

We will prove later on in Sect. \ref{sec1} the following proposition.
\begin{proposition}\label{p1.1}
 The two parameter group $\{U(t,s)\}_{t, s\in I}$  acting on $\coinf(\Sigma)\otimes \cc^{2}$ extends uniquely to a strongly continuous two parameter group $\{U(t,s)\}_{t,s\in I}$ such that $U(t,s): \cE_{s}\to \cE_{t}$  is unitary and $I^{2}\ni (t,s)\mapsto U(t,s)$ is strongly continuous (for the common topology of all the $\cE_{t}$).
 
 Denoting by $H(t)$ its infinitesimal generator we have:
\[
\begin{array}{rl}
i)&Dom H(t)= \Dom a(t)\oplus H^{1}(\Sigma),\\[2mm]
ii)&H(t)\hbox{ is selfadjoint on } \cE_{t}\hbox{ and }0\in \rho(H(t)).
\end{array} 
\]
\item 
\end{proposition}
\subsection{Vacuum states}
If $Q(t, \pe_{t}, x, \pe_{x})$ is a differential operator and $t_{0}\in I$ we denote
\begin{equation}
\label{defdept}
Q_{t_{0}}= Q(t_{0}, \pe_{t}, x , \pe_{x})
\end{equation}
the operator $Q$ with coefficients frozen at $t= t_{0}$.
In particular  $P_{t_{0}}= P(t_{0}, \pe_{t}, x, \pe_{x})$  is the Klein-Gordon operator $P$ with coefficients frozen at $t= t_{0}$. The associated Cauchy evolution is
$\e^{\i (t-s)H(t_{0})}$.  Since $P_{t_{0}}$ is invariant under time translations, and because of condition $(P)$, the quantized Klein-Gordon field for $P_{t_{0}}$
 admits a {\em vacuum state} $\omega^{\rm vac}_{t_{0}}$. Its covariances $\lambda^{\pm, {\rm vac}}_{t_{0}}$ are given by:
\beq\label{defdevac}
\lambda^{\pm, {\rm vac}}_{t_{0}}=  \pm q \circ \one_{\rr^{\pm}}(H(t_{0})),
\eeq
where $\one_{\rr^{\pm}}(H(t_{0}))$ are the spectral projections on $\rr^{\pm}$ for $H(t_{0})$, which are well defined by Prop. \ref{p1.1}.

\subsection{Adiabatic limits}\label{sec1.adiab}
 We fix  a compact interval $I\Subset \rr$ (for definiteness $I= [-1, 1]$) and consider  for $T\gg 1$ the Klein-Gordon operator
\beq\label{slow}
 P^{T}(t, \pe_{t}, x, \pe_{x})\defeq P(T^{-1}t, \pe_{t}, x,\pe_{x}),
\eeq
ie $h_{ij}(T^{-1}t, x)dx^{i}dx^{j}$, ${\rm A}_{i}(T^{-1}t, x)dx^{i}$, $V(T^{-1}t, x)$ and $m(T^{-1}t, x)$ are slowly varied from $t= -T$ to $t=T$.  Recalling the notation  in \eqref{defdept} we have \[
P^{T}_{\pm T}= P_{\pm 1}.
\]
 The associated Cauchy evolution $U_{T}(t,s)$ has  generator $H(T^{-1}t)$.

If $\lambda^{\pm}_{-1}$ are the covariances at time $t=-1$ of a state $\omega$ for the time-independent Klein-Gordon operator $P_{-1}$, we can investigate the existence of the adiabatic limit
\begin{equation}
\label{e.lim}
\lambda_{1}^{\rm ad}\eqdef \wlim_{T\to +\infty}U_{T}(-T, T)^{*}\lambda^{\pm}_{-1}U_{T}(-T, T)\hbox{ on }\coinf(\Sigma)\otimes \cc^{2}. 
\end{equation}
If the limits \eqref{e.lim} exist, then they are the time $t=1$ covariances of a quasi-free state $\omega^{\rm ad}$ for the time-independent Klein-Gordon operator  $P_{1}$.

We now state the main result of this paper.

\begin{theorem}\label{adiabatic-limit}
Let $\lambda^{\pm, {\rm vac}}_{-1}$ be the Cauchy surface covariances of the  vacuum state for the time-independent Klein-Gordon operator $P_{-1}$. Then 
the adiabatic limits
\[
 \wlim_{T\to +\infty}U_{T}(-T, T)^{*}\lambda^{\pm, {\rm vac}}_{-1}U_{T}(-T, T)\hbox{ exist on }\coinf(\Sigma)\otimes \cc^{2}
\]
and are the Cauchy surface covariances $\lambda^{\pm, {\rm vac}}_{1}$ of the  vacuum state for the time-independent Klein-Gordon operator $P_{1}$.
\end{theorem}
The proof, which follows directly from the adiabatic theorem Thm. \ref{theoadia},  is given in Subsect. \ref{sec2.3}.
\section{An adiabatic theorem for symplectic dynamics}\label{sec1}\init
In this section we prove a version of the adiabatic theorem of \cite{ASY} for a symplectic (instead of a unitary) dynamics. We will use the setup of Subsects. \ref{sec1.tot}, \ref{sec1.assumpt} although it is likely that the adiabatic theorem proved in Thm. \ref{theoadia} extends to a more general framework. A  natural situation   would be a two parameter (linear) symplectic flow generated by a time-dependent  quadratic Hamiltonian which is positive definite, corresponding to our condition $(P)$. Of course this positivity condition has to be supplemented by abstract versions of $(H)$, $(D)$, implying for example that the energy norms are locally uniformly equivalent to some  reference Hilbert  norm.
We assume  hence hypotheses $(H)$, $(P)$, $(D)$. We start by proving Prop. \ref{p1.1}.

\subsection{Proof of Prop. \ref{p1.1}}
 On $\coinf(\Sigma)\otimes \cc^{2}$ we have:
\[
\p_{t}U(t,s)= \i H(t)U(t,s), \ \p_{s}U(t,s)= -\i U(t,s)H(s),
\]
for 
\beq\label{e1.00}
H(t)= \mat{V(t)}{\one}{a(t)}{V(t)}.
\eeq
It is  convenient to  set:
\[
\hat{\rho}_{s}\phi(x)=\col{\phi(s,x)}{\i^{-1}\p_{t}\phi(s,x)}=: g
\] so that $\hat{\rho}_{s}\phi= S(s)\rho_{s}\phi$, $S(s)= \mat{\one}{0}{V(s)}{\one}$.  The associated evolution is
\beq\label{e1.-2}
W(t,s)= S(t)U(t,s)S^{-1}(s), 
\eeq
with generator 
\[
\begin{array}{rl}
K(t)=& S(t)H(t)S^{-1}(t)- \i \p_{t}S(t)S^{-1}(t)\\[2mm]
=& \mat{0}{\one}{p(t)- \i \p_{t}V(t)}{2V(t)}, 
\end{array}
\]
where  we recall that $p(t)= a(t)-V^{2}(t)$. We set
 \[
F_{t}(g, g)= (g_{1}| g_{1})+ (g_{0}| p(t)g_{0}).
\]
 Again the completion of $\coinf(\Sigma)\otimes \cc^{2}$ for $F_{t}^{\12}$ equals $\cE_{t}$.
We obtain that if $g(t)= W(t,s)g$, $g\in \coinf(\Sigma)\otimes \cc^{2}$:
\beq\label{e1.0}
\begin{array}{rl}
&\p_{t}F_{t}(g(t), g(t))=(g_{1}(t)|\p_{t}V(t)g_{0}(t))\\[2mm]
+ &(g_{0}(t)| \p_{t}V(t)g_{1}(t))+ (g_{0}(t)| \p_{t}p(t)g_{0}(t)).
\end{array}
\eeq
Using $(H)$ and $(P)$ we obtain  that for $t\in I$ one has:
\[
| (g_{0}(t)| \p_{t}p(t) g_{0}(t))|\leq C_{I}F_{t}(g_{t}, g_{t}), 
\]
which using also $(H)$ for the other terms in the rhs of \eqref{e1.0} yields
\[
|\p_{t}F_{t}(g_{t}, g_{t})|\leq C_{I}F_{t}(g_{t}, g_{t}), \ t\in I.
\]
By Gronwall's inequality this implies that for any $I\Subset \rr$ we have:
\[
\sup_{t, s\in I}\|W(t,s)\|_{B(\cE_{t})}\leq C_{I}.
\]
Since $W(t,s)$ is strongly continuous on  the dense subspace $\coinf(\Sigma)\otimes \cc^{2}$ it is strongly continuous on $\cE_{t}$. By \eqref{e1.-2} the same is true for $U(t,s)$.

The operator $K(t)$ preserves $\coinf(\Sigma)\otimes \cc^{2}$ and is bounded from $\cE_{t}$ to $\cE_{t}^{*}= L^{2}(\Sigma)\oplus H^{-1}(\Sigma)$.  Its domain as the infinitesimal generator of $W(t,s)$ is 
\[
\Dom K(t)= \{g\in \cE_{t}\ : \ K(t)g\in \cE_{t}\}= \Dom a(t)\oplus H^{1}(\Sigma),
\]
by direct inspection, using that $\Dom a(t)= \{u\in H^{1}(\Sigma)\ : \ a(t)u\in L^{2}(\Sigma)\}$. 

Note that using $(Hiv)$ we obtain that $V(t): H^{1}(\Sigma)\to H^{1}(\Sigma)$ hence  $S(t)$ is an isomorphism of both $\cE_{t}$ and   of $\Dom a(t)\oplus H^{1}(\Sigma)$. Therefore the domain of $H(t)$ as infinitesimal generator of $U(t,s)$ equals $S(t)^{-1}\Dom K(t)= \Dom a(t)\oplus H^{1}(\Sigma)$.

We now study the operator $H(t)$.  Let us set  
\[
L(t)= S(t)H(t)S(t)^{-1}= \mat{0}{\one}{p(t)}{2V(t)}.
\] From $(P)$ we know that $0\in \rho(p(t))$ hence $0\in \rho(L(t))$ by \cite[Prop. 5.3]{GGH}. Using then \cite[Thm. 5.4]{GGH} we obtain that $L(t)$ is selfadjoint on $\cE_{t}$.  This implies that $H(t)$ is selfadjoint on $\cE_{t}$ with $0\in \rho(H(t))$. \qed
\subsection{Smoothness of spectral projections}\label{sec1.3}
Since by Prop. \ref{p1.1} $H(t)$ is selfadjoint on $\cE_{t}$ we can define the spectral projection
\[
P(t)= \one_{\rr^{+}}(H(t))\in B(\cE_{t}).
\]
Moreover since $0\in \rho(H(t))$,  for each $I\Subset \rr$ there exist  $\chi\in \cinf(\rr)$, $\chi\equiv 1$ near $+\infty$ such that  $P(t)= \chi(H(t))$ for $t\in I$. In this subsection we examine the smoothness of $P(t)$ w.r.t. $t$. 
\subsubsection{Almost analytic extensions and functional calculus}
Let us set 
\[
S^{\rho}(\rr)=\{f\in \cinf(\rr) : \p_{\lambda}^{n}f(\lambda)\in O(\langle \lambda\rangle^{\rho-n}),  n\in \nn\}, \ \rho\in \rr.
\]
We equip $S^{\rho}(\rr)$ with the semi-norms $\|f\|_{\rho, n}= \sup_{\lambda\in \rr}|\langle \lambda \rangle^{\rho-n}\p_{\lambda}^{n}f(\lambda)|$.

For $f\in S^{\rho}(\rr)$ we denote by $\tilde{f}\in \cinf(\cc)$ an {\em almost analytic extension} of $f$ satisfying:
\beq\label{alamo}
\begin{array}{rl}
i)& \tilde{f}\traa{\rr}= f, \\[2mm]
ii)&\supp \tilde{f}\subset \{|{\rm Im}z|\leq C |{\rm Re}z|\},\\[2mm]
iii)&|\p_{\bar{z}}\tilde{f}(z)|\in O(\langle z\rangle)^{\rho-1-k}| {\rm Im}z|^{k}, \ \forall k\in \nn, 
\end{array}
\eeq
see for example \cite[Prop. C.2.2]{DG97} for a construction of $\tilde{f}$.  
If $H$ is selfadjoint on a Hilbert space $\cH$  we have the bounds:
\begin{equation}
\label{titu}
\begin{array}{rl}
i)& \| (H-z)^{-1}\|\leq |{\rm Im}z|^{-1},\\[2mm]
ii)&\|(H+\i)(H-z)^{-1}\| \leq c\langle z\rangle |{\rm Im}z|^{-1} \hbox{ for }|{\rm Im}z|\leq C |{\rm Re}z|.
\end{array}
\end{equation}
 and $f\in S^{\rho}(\rr)$,  for $\rho<0$, then one has
\beq\label{e0.03}
f(H)=\frac{1}{2\i\pi}\int_{\cc}\p_{\bar{z}}\tilde{f}(z)(z-H)^{-1}dz\wedge d\bar{z},
\eeq
the integral being norm convergent, using \eqref{titu} {\it i)}. 

Let us now explain how to extend \eqref{e0.03} to the case $\rho\geq 0$.
Let us fix $\chi\in \coinf(\rr)$ with $\chi= 1$ near $0$ and $\tilde{\chi}\in \coinf(\cc)$ an almost analytic extension of $\chi$. We set $\chi_{R}(x)= \chi(R^{-1}x)$ and $\tilde{\chi}_{R}(z)= \tilde{\chi}(R^{-1}z)$, which is an almost analytic extension of $\chi_{R}$.

For $f\in S^{\rho}(\rr)$ and $\rho\geq 0$,   we set $f_{R}(\lambda)= f(\lambda)\chi_{R}(\lambda)$. We have: 
\beq
\label{alternat.0}
\begin{array}{l}
\{f_{R}\}_{R\geq 1}\hbox{ is  bounded in }S^{\rho}(\rr), \\[2mm]
R^{\rho'-\rho} \{(f_{R}- f)\}_{R\geq 1}\hbox{  is bounded in }S^{\rho'}(\rr), \ \forall \rho'>\rho.
\end{array}
\eeq
Let  us set $\tilde{f}_{R}= \tilde{f}\tilde{\chi}_{R}$,  which is an almost analytic extension of $f_{R}$.  The following properties of $\tilde{f}_{R}$ follow from \eqref{alternat.0} and the construction of $\tilde{f}, \tilde{\chi}$ in \cite{DG97}:
\begin{equation}
\label{alt.e2}
\begin{array}{rl}
i)& \tilde{f}_{R}\traa{\rr}= f_{R}, \\[2mm]
ii)&\supp \tilde{f}_{R}\subset \{|{\rm Im}z|\leq C |{\rm Re}z|\}\cap \{|z|\leq CR\},\\[2mm]
iii)&|\p_{\bar{z}}\tilde{f}_{R}(z)|\in O(\langle z\rangle)^{\rho-1-k}| {\rm Im}z|^{k}, \ \forall k\in \nn, \hbox{ uniformly for }R\geq 1,\\[2mm]
iv)&|\p_{\bar{z}}(\tilde{f}(z)- \tilde{f}_{R}(z))|\in O(\langle z\rangle)^{\rho'-1-k}R^{\rho-\rho'}| {\rm Im}z|^{k}, \ \forall k\in \nn, \ \rho'>\rho.
\end{array}
\end{equation}
Since $f(H)= \slim_{R\to +\infty}f_{R}(H)$ in $B(\Dom |H|^{\rho}, \cH)$ we have:
\beq\label{alternat.1}
f(H)=\slim_{R\to +\infty}\frac{1}{2\i\pi}\int_{\cc}\p_{\bar{z}}(\tilde{f}_{R})(z)(z-H)^{-1}dz\wedge d\bar{z},\hbox{ in }B(\Dom|H|^{\rho}, \cH).
\eeq

\begin{lemma}\label{2.1}
 Assume $(H)$, $(P)$ and $(D)$. Let $\epsilon(t)= a(t)^{\12}$. Then for $k=1,2$:
 \[
\begin{array}{rl}
i)&\epsilon(t)(\epsilon^{-1})^{(k)}(t) \hbox{  is bounded on }\cH,\hbox{ locally  uniformly in  }t,\\[2mm]
ii)&\epsilon^{-1}(t)\epsilon^{(k)}(t) \hbox{ is bounded on }\cH,\hbox{ locally  uniformly in  }t.
\end{array}
\]\end{lemma}
\proof  Note that by duality and interpolation $(Di)$ implies that $\epsilon^{-1}(t)a'(t)\epsilon^{-1}(t)$ is bounded on $\cH$ locally uniformly in $t$. Let us first prove $i)$. We have $\epsilon^{-1}(t)= \pi^{-1}\int_{0}^{+\infty}\lambda^{-\12}(a(t)+ \lambda)^{-1}d\lambda$, hence:
\[
\begin{array}{rl}
\epsilon(t)(\epsilon^{-1})'(t)=&-\pi^{-1}\int_{0}^{+\infty}\lambda^{-\12}a(t)^{\12}(a(t)+ \lambda)^{-1}a'(t)(a(t)+ \lambda)^{-1}d\lambda\\[2mm]
=&-\pi^{-1}\int_{0}^{+\infty}\lambda^{-\12}a(t)^{\12}a'(t)(a(t)+ \lambda)^{-2}d\lambda\\[2mm]
&+\pi^{-1}\int_{0}^{+\infty}\lambda^{-\12}a(t)^{\12}(a(t)+ \lambda)^{-1}[a(t), a'(t)](a(t)+\lambda)^{-2}d\lambda.
\end{array}
\]
The first term equals $a(t)^{\12}a'(t)a(t)^{-3/2}$ which is bounded by $(Dii)$. We write the second term as
\[
\pi^{-1}\int_{0}^{+\infty}\lambda^{-\12}a(t)(a(t)+ \lambda)^{-1}a(t)^{-\12}[a(t), a'(t)]a(t)^{-1}(a(t)+\lambda)^{-2}a(t)d\lambda.
\]
The integral is norm convergent using $(Diii)$ since $a(t)(a(t)+ \lambda)^{-1}\in O(1)$. Therefore $\epsilon(t)(\epsilon^{-1})'(t)$ is bounded on $L^{2}(\Sigma)$ which proves $i)$. To prove $ii)$ we write $\epsilon(t)= \epsilon^{-1}(t)a(t)$ hence
\[
\epsilon'(t)\epsilon^{-1}(t)= \epsilon^{-1}(t)a'(t)\epsilon^{-1}(t)+ (\epsilon^{-1})'(t)a(t)\epsilon^{-1}(t)
\] 
which is bounded on $L^{2}(\Sigma)$ by $i)$ and $(Di)$.  Using the same argument we prove the estimates for second derivatives. \qed
\begin{proposition}\label{2.2}
 Assume $(H)$, $(P)$ and $(D)$. Let $P(t):= \one_{\rr^{+}}(H(t))$. Then $\rr\ni t\mapsto P(t)\in B(\cE_{t})$ is strongly $C^{2}$ and $P^{(k)}(t)$ is bounded on $\cE_{t}$ locally uniformly in $t$ for $k=1,2$.
 \end{proposition}
\proof 
 In the sequel we write $A(t)\in O(1)$ if $\|A(t)\|_{B(\cE_{t})}\in O(1)$ for $t\in I$. We set $H_{0}(t)= \mat{0}{\one}{a(t)}{0}$, $W(t):= H(t)- H_{0}(t)\in O(1)$, using $(H iii))$. 
  Since $0\in \rho(H_{(0)}(t))$ we have $\one_{\rr^{+}}(H_{(0)}(t))= f(H_{(0)}(t))$, for $t\in I$, for some $f\in S^{0}(\rr)$, $f=1$ near $+\infty$, $f=0$ near $-\infty$.  
 We have by \eqref{alternat.1}:
 \[
 f(H(t))= f(H_{0}(t))+ R(t),
 \]
 for
 \[
 R(t)= \slim_{R\to +\infty}\frac{1}{2\i \pi}\int_{\cc}\p_{\bar z}\tilde{f}_{R}(z)(z- H(t))^{-1}W(t)(z- H_{0}(t))^{-1}dz\wedge d\bar{z}, \hbox{ in }\cE_{t}.
 \]
Using \eqref{alt.e2} $iv)$  we then obtain
\[
R(t)= \frac{1}{2\i \pi}\int_{\cc}\p_{\bar z}\tilde{f}(z)(z- H(t))^{-1}W(t)(z- H_{0}(t))^{-1}dz\wedge d\bar{z},
\]
the integral being norm convergent on $B(\cE_{t})$.

We have by an easy computation $f(H_{0}(t))= \one_{\rr^{+}}(H_{0}(t))= \12\mat{\one}{\epsilon(t)^{-1}}{\epsilon(t)}{\one}$,
which using Lemma \ref{2.1} implies that $ f(H_{0}(t))'$ is uniformly bounded on $\cE_{t}$ for $t\in I$.  Next we compute:
\[\begin{array}{l}
R'(t)= \frac{\i}{2\pi}\int_{\cc}\p_{\bar z}\tilde{f}(z)(z- H(t))^{-1}W'(t)(z- H_{0}(t))^{-1}dz\wedge d\bar{z}\\[2mm]
+ \frac{\i}{2\pi}\int_{\cc}\p_{\bar z}\tilde{f}(z)(z- H(t))^{-1}H'(t)(z- H(t))^{-1}W(t)(z- H_{0}(t))^{-1}dz\wedge d\bar{z}\\[2mm]
+ \frac{\i}{2\pi}\int_{\cc}\p_{\bar z}\tilde{f}(z)(z- H(t))^{-1}W(t)(z- H_{0}(t))^{-1}H'_{0}(t)(z- H_{0}(t))^{-1}dz\wedge d\bar{z}.
\end{array}
\]
From  $(H)$ we see easily that $W'(t)\in O(1)$ and  from $(Di)$  that $H'_{0}(t)(H_{0}(t)+ \i)^{-1}$, $H'(t)(H(t)+ \i)^{-1}\in O(1)$.
Using also \eqref{titu} we obtain that the integrands in the rhs above  are bounded by either $|{\rm Im}z|^{-2}$ or by $\langle z\rangle |{\rm Im}z|^{-3}$, uniformly for $t\in I$. Since $\p_{\bar z}\tilde{f}\in O(\langle z\rangle^{-1- k})|{\rm Im}z|^{k}$ we obtain that $R'(t)\in O(1)$. We use the same argument to estimate $P''(t)$.  \qed

 \qed 

The following lemma is a version of \cite[Lemma 2.5]{ASY}, where the case when $P(t)$ is the spectral projection on a bounded interval was considered.
\begin{lemma}\label{2.3}
 Let $I\ni t\mapsto H(t)$ be a map with values in selfadjoint operators  on a Hilbert space $\cH$ and  $I\ni t\mapsto X(t)\in B(\cH)$ be strongly $C^{1}$.  Assume that $\cD= \Dom H(t)$ is independent on $t$ and that:
 \[
 H(t): \cD\to \cH\hbox{ is strongly  differentiable}, \ [-\alpha, \alpha]\subset \rho(H(t))\hbox{ for }t\in I.
\]  
Let $P(t)= \one_{\rr^{+}}(H(t))$ and let us fix $f\in S^{0}(\rr)$ such that $f(\lambda)= \one_{\rr^{+}}(\lambda)$   $\rr\backslash[-\alpha, \alpha]$   and $\tilde{f}$  an almost analytic extension of $f$ satisfying \eqref{alamo}.
Then  the integral
\beq\label{e12.00}
\tilde{X}(t)\defeq -\frac{1}{2\i\pi}\int_{\cc}\p_{\bar z} \tilde{f}(z)(z-H(t))^{-1}X(t)(z-H(t))^{-1}dz\wedge d\bar{z},
\eeq
is norm convergent in $B(\cH)$.

The map $I\ni t\mapsto \tilde{X}(t)\in B(\cH)$  is strongly $C^{1}$ and
\[
[P(t), X(t)]= [\tilde{X}(t), H(t)], \hbox{ as quadratic forms on }\cD.
\] 
\end{lemma}
\proof  
We first fix $t\in I$ and omit the parameter $t$ for simplicity of notation. 
  Using \eqref{alt.e2} we obtain
\[
P= \slim_{R\to +\infty}\frac{1}{2\i \pi}\int_{\cc}\p_{\bar z}\tilde{f}_{R}(z)(z-H)^{-1}dz\wedge d\bar{z}, \hbox{ in }B(\cH),
\]
hence:
\[
[P, X]=\slim_{R\to +\infty}\frac{1}{2\i\pi}\int_{\cc}\frac{\p \tilde{f}_{R}}{\p \bar{z}}(z)[(z-H)^{-1}, X]dz\wedge d\bar{z}, \hbox{ in }B(\cH).
\]
 We recall that $\cD= \Dom H(t)$ is independent on $t$ and denote by $\cD'$ its topological dual. 
Since $[(z-H)^{-1}, X]= [H, (z-H)^{-1}X(z-H)^{-1}]$ on  $B(\cD, \cD')$, we obtain
\beq\label{alt.e3}
[P, X]=\slim_{R\to +\infty}\frac{1}{2\i\pi}\int_{\cc}\p_{\bar z}\tilde{f}_{R}(z)[H, (z-H)^{-1}X(z-H)^{-1}]dz\wedge d\bar{z}\hbox{ in }B(\cD, \cD').
\eeq
 Using \eqref{alt.e2} $iv)$  we can compute the strong limit in the rhs of \eqref{alt.e3} and obtain $[P, X]= [H, \tilde{X}]$ for
\[
\tilde{X}= -\frac{1}{2\i \pi}\int\p_{\bar z}\tilde{f}(z)(z-H)^{-1}X(z-H)^{-1}dz\wedge d\bar{z}.
\]
  It remains to check that $t\mapsto \tilde{X}(t)$ is strongly $C^{1}$.  This follows from differentiating in $t$ the rhs of \eqref{e12.00}, using $\frac{d}{dt}(z- H(t))^{-1}= (z-H(t))^{-1}H'(t)(z-H(t))^{-1}$. The details are left to the reader. \qed

\subsection{Adiabatic evolution}\label{sec2.2}
We recall from Subsect. \ref{sec1.adiab} that $U_{T}(t,s)$ is the Cauchy evolution associated to the Klein-Gordon operator $P(T^{-1}t, \pe_{t}, x, \pe_{x})$.
Repeating the computations in the proof of Prop. \ref{p1.1}  we obtain  for $f(t)= \tilde{U}_{T}(t,s)f$ that 
\[
\p_{t}E_{t}(f(t),f(t))\leq CT^{-1} E_{t}(f(t),f(t)), t\in [-T, T],
\]
hence by Gronwall's inequality
\[
\| U_{T}(t,s)f)\|_{\cE_{t}}\leq C\|f\|_{\cE_{s}}, \ t,s\in [-T, T],
\]
where  we recall that the norm $\|\cdot\|_{\cE_{t}}$ is defined in Def. \ref{defoen}.
\[
\cE_{t}= H^{1}(\Sigma)\oplus L^{2}(\Sigma),
\]
equipped with the norm $E_{t}(f, f)^{\12}$ introduced in \eqref{e1.01}.  This implies that
\[
\|U_{T}(t,s)\|_{B(\cE)}\leq C,  \ t,s\in [-T, T].
\]
We set 
\[
\hat{U}_{T}(t,s)=U_{T}(Tt, Ts), \ t,s\in[-1, 1]. 
\]
whose generator is $TH(t)$. We obtain that:
\begin{equation}
\label{e2.3}
\|\hat{U}_{T}(t,s)\|_{B(\cE)}\leq C, \ t,s\in [-1, 1], \ T\geq 1.
\end{equation}

\def\ad{{\rm ad}}\def\Uad{\hat{U}_{T}^{\ad}}
We set 
\[
H^{\ad}_{T}(t):= H(t)+ \i T^{-1}[P(t), P'(t)],
\]
where $P(t)= \one_{\rr^{+}}(H(t))$. Since $[-1, 1]\ni t\mapsto P(t), P'(t)\in B(\cE_{t})$ are strongly continuous by Prop. \ref{2.2}, the evolution group $\hat{U}_{T}^{\ad}(t,s)$ with generator $TH^{\ad}_{T}(t)$ can be constructed by setting:
\[
\hat{U}_{T}^{\ad}(t,s)=: \hat{U}_{T}(t,0)Z_{T}(t,s)\hat{U}_{T}(0,s),
\]
where 
\[
\begin{array}{l}
\p_{t}Z_{T}(t,s)=  K_{T}(t)Z_{T}(t,s), \ Z_{T}(s,s)=\one, \\[2mm]
 K_{T}(t)= \hat{U}_{T}(0,t)  [P'(t),P(t)]\hat{U}_{T}(t, 0)\in B(\cE_{t}).
\end{array}
\]
By a standard argument (see eg \cite[Lemma 2.3]{ASY}) one obtains that:
\begin{equation}
\label{e2.3b}
P(t)\hat{U}^{\ad}_{T}(t,s)= \hat{U}^{\ad}_{T}(t,s)P(s), \ t,s\in[-1, 1].
\end{equation}
In fact it suffices to differentiate both terms in $t$, after acting on a vector in $\Dom H(t)$. Moreover from \eqref{e2.3} we obtain
\begin{equation}
\label{e2.4}
\|\hat{U}^{\ad}_{T}(t,s)\|_{B(\cE)}\leq C, \ t,s\in [-1, 1], \ T\geq 1.\end{equation}

\def\bP{\bar{P}}\def\dP{P'}
\subsection{Adiabatic theorem}\label{sec2.3}
We now state a version of the adiabatic theorem which is sufficient for our purposes.
\begin{theorem}\label{theoadia}
 Assume $(H)$, $(P)$ and $(D)$. Then  there exists $C>0$ such that:
 \[
\| \hat{U}_{T}(t,s)- \hat{U}^{\ad}_{T}(t,s)\|_{B(\cE)}\leq CT^{-1}, \ t,s\in [-1, 1].
\]
\end{theorem}
 The theorem can be proved by repeating the arguments in the proof of \cite[Thm. 2.4]{ASY}. For the reader's convenience we will sketch its main steps.

We  often remove the time variable for simplicity of notation.
We set $\bP= 1-P$, and denote by  $\tilde{X}$ the operator constructed in Lemma \ref{2.3} for some  strongly $C^{1}$ map $t\mapsto X(t)$. From $P^{2}=P$ we obtain $PP'+ P'P= P'$ hence 
\beq\label{e3.9}
[P, \dP]= 2P\dP-\dP= 2\dP\bP- \dP.
\eeq
It follows that:
\beq\label{e3.10}
\begin{array}{rl}
&\bP XP=\bP[X, P]P= \bP[H, \tilde{X}]P= \bP[H^{\ad}, \tilde{X}]P)- \i T^{-1}\bP[[P, \dP], \tilde{X}]P\\[2mm]
=& \bP[H^{\ad}, \tilde{X}]P+\i T^{-1}\bP[\dP, \tilde{X}]P.
\end{array}
\eeq
\begin{lemma}\label{toto}
 Assume that $[-1, 1]\ni t\mapsto X(t), Y(t)\in B(\cE)$ are strongly $C^{1}$. Then for $t,s\in[-1, 1]$:
 \[
\begin{array}{rl}
&\int_{s}^{t}\bP(s)\Uad(s, t_{1})X(t_{1})\Uad(t_{1},s)P(s)Y(t_{1})dt_{1}\\[2mm]
=&-\i T^{-1}\left[\bP(s)\Uad(s, t_{1})\tilde{X}(t_{1})\Uad(t_{1}, s)P(s)Y(t_{1})\right]_{s}^{t}\\[2mm]
&+\i T^{-1}\int_{s}^{t}\bP(s)\Uad(s, t_{1})\tilde{X}'(t_{1})\Uad(t_{1},s)P(s)Y(t_{1})dt_{1}\\[2mm]
&+\i T^{-1}\int_{s}^{t}\bP(s)\Uad(s, t_{1})\tilde{X}(t_{1})\Uad(t_{1},s)P(s)Y'(t_{1})dt_{1}\\[2mm]
&+\i T^{-1}\int_{s}^{t}\bP(s)\Uad(s, t_{1})[P'(t_{1}),\tilde{X}(t_{1})]\Uad(t_{1},s)P(s)Y(t_{1})dt_{1}.
\end{array}
\]
In particular we have:
\beq\label{e3.12}
\|\int_{s}^{t}\bP(s)\Uad(s, t_{1})X(t_{1})\Uad(t_{1},s)P(s)Y(t_{1})dt_{1}\|_{B(\cE)}\leq C T^{-1},
\eeq
where the constant $C$ depends only on $\sup_{t\in [-1, 1]}\|X^{(k)}(t)\|+ \| Y^{(k)}(t)\| + \| P^{(k)}(t)\|$, $k= 0,1$.
\end{lemma}
\proof
From \eqref{e3.10} we obtain:
\[
\begin{array}{l}
\bP(s)\Uad(s,t)X(t)\Uad(t,s)P(s)=\Uad(s,t)\bP(t)X(t)P(t)\Uad(t,s)\\[2mm]
=\Uad(s,t)\bP(t)[H^{\ad}(t), \tilde{X}(t)]P(t)\Uad(t,s)+\i T^{-1}\Uad(s,t)\bP(t)[\dP(t), \tilde{X}(t)]P(t)\Uad(t,s)\\[2mm]
= \i T^{-1}\bP(s)\left(-\p_{t}(\Uad(s,t)\tilde{X}(t)\Uad(t,s))+ \Uad(s,t)(\tilde{X}'(t)+ [P'(t), \tilde{X}(t)])\Uad(t,s)\right)P(s).
\end{array}
\]
The lemma follows by integration by parts. \qed

{\bf Proof of Thm. \ref{theoadia}.}

We set for fixed $s\in [-1, 1]$ $\Omega_{T}(t,s)\defeq \Uad(s,t)\hat{U}_{T}(t,s)$, so that
\beq\label{e3.120}
\Omega_{T}(t,s)= \one + \int_{s}^{t}R_{T}(t_{1})\Omega_{T}(t_{1}, s)dt_{1},
\eeq
\[
R_{T}(t)= \Uad(s,t)[P(t), \dP(t)]\Uad(t,s).
\]
From \eqref{e2.3b}, \eqref{e3.9} we have:
\[
\begin{array}{l}
\bP(s)R_{T}(t)= \Uad(s,t)\bP(t)[P(t), \dP(t)]\Uad(t,s)\\[2mm]
= \Uad(s,t)\bP(t)[P(t), \dP(t)]P(t)\Uad(t,s)\\[2mm]
= -\bP(s)\Uad(s,t)\dP(t)\Uad(t,s)P(s).
\end{array}
\]
Applying Lemma \ref{toto} to $X(t)= \dP(t)$,  $Y(t)= \Omega_{T}(t,s)$ we obtain from \eqref{e3.12}
\[
\|\int_{s}^{t}\bP(s) R_{T}(t_{1})\Omega_{T}(t_{1},s)dt_{1}\|\leq CT^{-1}.
\]
Exchanging the role of $P$ and $\bP$ we also obtain
\[
\|\int_{s}^{t}P(s) R_{T}(t_{1})\Omega_{T}(t_{1},s)dt_{1}\|\leq CT^{-1},
\]
hence
\[
\|\Omega_{T}(t,s)- \one\|_{B(\cE)}\leq CT^{-1}, \ t,s\in [-1, 1].
\]
This implies that
\[
\| \hat{U}_{T}(t,s)- \hat{U}^{\ad}_{T}(t,s)\|_{B(\cE)}\leq CT^{-1}, \ t,s\in [-1, 1]
\]
and completes the proof. \qed

\medskip

{\bf Proof of Thm. \ref{adiabatic-limit}.}
 Since $U_{T}(t,s)$ is symplectic, we have $U_{T}(-T, T)^{*}qU_{T}(-T, T)=q$, hence  using \eqref{defdevac}:
\[
\begin{array}{rl}
&U_{T}(-T, T)^{*}\lambda^{\pm, \rm vac}_{-1}U_{T}(-T, T)= q U_{T}(T, -T)\one_{\rr^{\pm}}(H(-1))U_{T}(-T, T)\\[2mm]
=&q \hat{U}_{T}(1, -1)\one_{\rr^{\pm}}(H(-1))\hat{U}_{T}(-1, 1)=q \hat{U}^{\ad}_{T}(1, -1)\one_{\rr^{\pm}}(H(-1))\hat{U}^{\ad}_{T}(-1, 1)+ O(T^{-1})\\[2mm]
=&q\one_{\rr^{\pm}}(H(1))+ O(T^{-1})= \lambda^{\pm, {\rm vac}}_{1}+ O(T^{-1}),
\end{array}
\]
using Thm. \ref{theoadia} and \eqref{e2.3b} (remember that $P(t)=\one_{\rr^{+}}(H(t))$). \qed
\section{Further results in the separable case}\label{sec4}\init
We consider now a simpler version of the setup in Subsects. \ref{sec1.tot}, \ref{sec1.assumpt} where $A= 0$, $h_{t}= h$ is independent on $t$ and $m(t, x)=m^{2}(t)$.  We assume
\[
(HC)\ \Sigma\hbox{ is non compact, }  \sigma(-\Delta_{h})= [0, +\infty[\hbox{ is purely absolutely continuous}.
\] 
The Klein-Gordon operator 
takes the form:
\[
\tilde{P}= P= \pe_{t}^{2}- \Delta_{h}+ \chi(t)m^{2}\eqdef\pe_{t}^{2}+ a(t),
\]
 where $a(t)$ commutes with $-\Delta_{h}$. It follows that the Klein-Gordon equation $P\phi=0$ can be reduced to a family of $1-d$ Schr\"{o}dinger equations:
\beq\label{decompo}
\p_{t}^{2}\phi+ m^{2}\chi(t)\phi+ \epsilon^{2}\phi=0, 
\eeq
where $\epsilon= (-\Delta_{h})^{\12}$, if one introduces a spectral decomposition of $\epsilon$. This is known in the physics literature as the {\em mode decomposition} method.

We show   in Thm. \ref{th4.2}    that the conclusion  of Thm. \ref{adiabatic-limit}  still holds when the  initial or final mass $m(\mp 1)$ vanishes, ie when  the stability condition $(P)$ is violated.
 
 We next consider the adiabatic limit of an initial {\em thermal state} at temperature $\beta^{-1}$, and show in Thm. \ref{th4.1} that its adiabatic limit is {\em not}  a thermal state (unless the initial and final masses are the same).
 
 Finally we consider the adiabatic limits of (infrared regular) Hadamard states and show  in Thm. \ref{th4.3} that their adiabatic limits are again Hadamard states.

\subsection{Energy estimates}\label{sec4.1}
We  will set:
 \[
\epsilon\defeq (-\Delta_{h})^{\12}, \ \epsilon_{t}= \epsilon(t)\defeq a(t)^{\12}, \ m_{t}= m(t).
\]
 We set as in Subsect. \ref{sec2.2}:
\[
\hat{U}_{T}(t,s)= \Texp(\i T\int_{s}^{t}H(\sigma)d\sigma), \ t,s\in [-1, 1],
\]
where now: 
\[
H(t)= \mat{0}{\one}{a(t)}{0}.
\]
We will consider the following three cases:
\[
\begin{array}{rl}
A:& m(t)>0, \ t\in[-1, 1],\\[2mm]
B:&m_{-1}=0 , \ m(t) \hbox{ strictly increasing},\\[2mm]
C:& m_{1}=0, \ m(t)\hbox{ strictly decreasing}.
\end{array}
\]
Conditions $(H)$ and $(D)$ are always satisfied but condition $(P)$ is not satisfied in cases $B$ and $C$. 
\subsubsection{Modified energy spaces}
We set  (recall that $\cH= L^{2}(\Sigma, dVol_{h})$):
\def\cC{\mathcal{C}}
\def\cA{\mathcal{A}}
\def\cB{\mathcal{B}}
\begin{equation}
\label{e3.200}
\begin{array}{l}
\cA:= \langle \epsilon\rangle^{-\12}\cH\oplus \langle \epsilon\rangle^{\12}\cH,\\[2mm]
\cB_{t}:= \epsilon_{t}^{-1}\epsilon^{\12}\cH \oplus \epsilon^{\12}\cH,\\[2mm]
\cC_{t}:= \langle \epsilon\rangle^{-\12}\cH \oplus \epsilon_{t}\langle \epsilon\rangle^{-\12}\cH.
\end{array}
\end{equation}
which are well defined since $\Ker \epsilon=\{0\}$. We recall from Subsect. \ref{sec0.not} that   if
\beq\label{defdes}
\cS=\{f\in\cH\otimes \cc^{2}: f= \one_{[\delta, R]}(\epsilon)f, \ R, \delta>0\},
\eeq
 $\cA$, $\cB_{t}$ $\cC_{t}$ are the completion of $\cS$ for  the norms:
\begin{equation}
\label{e3.12bb}
\begin{array}{l}
\|f\|_{\cA}^{2}\defeq\| \langle \epsilon\rangle^{\12}f_{0}\|^{2}_{\cH}\oplus \| \langle \epsilon\rangle^{-\12}f_{1}\|_{\cH}^{2},\\[2mm]
\|f\|_{\cB_{t}}^{2}\defeq \|\epsilon^{-\12}\epsilon_{t}f_{0}\|_{\cH}^{2}+ \|\epsilon^{-\12}f_{1}\|_{\cH}^{2},
\\[2mm]
\|f\|_{\cC_{t}}^{2}\defeq \|\langle \epsilon\rangle^{\12}f_{0}\|_{\cH}^{2}+ \|\langle \epsilon\rangle^{\12}\epsilon_{t}^{-1}f_{1}\|_{\cH}^{2}.
\end{array}
\end{equation}
\subsubsection{Energy estimates}
\begin{lemma}\label{l3.1}
The following estimates hold for $t\leq s$, $t,s\in [-1, 1]$:
\begin{equation}
\label{e3.99}
\begin{array}{rl}
\hbox{ case } A:\ \|\hat{U}_{T}(t,s)\|_{B(\cA)}\leq C,\\[2mm]
\hbox{ case } B:\ \|\hat{U}_{T}(t,s)\|_{B(\cB_{s}, \cB_{t})}\leq C,\\[2mm]
\hbox{ case } C:\ \|\hat{U}_{T}(t,s)\|_{B(\cC_{s}, \cC_{t})}\leq C.
\end{array}
\end{equation}
\end{lemma}
\proof 
The estimate for case $A$ follows from \eqref{e2.3}, using that $\hat{U}_{T}(t,s)$ commutes with $\langle \epsilon\rangle^{\12}$.
In case $B$, if  $f(t)= \hat{U}_{T}(t,s)f$ since  $a'(t)=2m(t)m'(t)\geq 0$, we obtain that $\frac{\d}{\d t}\|f(t)\|^{2}_{\cB_{t}}\geq 0$ which implies the desired estimate.  The same argument  can be used for case $C$. \qed
\subsection{Adiabatic limit of sesquilinear forms}\label{sec4.2}
We identify sesquilinear forms on $\cA$, $\cB_{t}$ or $\cC_{t}$ with linear operators. In fact the canonical scalar product on $\cH\otimes \cc^{2}$ allows to identify $\cA^{*}$ with $\langle \epsilon\rangle^{\12}\cH\oplus \langle \epsilon\rangle^{-\12}\cH$, $\cB_{t}^{*}$ with $\epsilon^{\12}\epsilon_{t}\cH\oplus \epsilon^{-\12}\cH$, $\cC_{t}$ with $\langle \epsilon\rangle^{\12}\cH\oplus \langle \epsilon\rangle^{\12}\epsilon_{t}^{-1}\cH$. In this way we 
will identify a sesquilinear form $\lambda$  with a linear operator, still denoted by $\lambda$, by
\[
\bar{f}\dito \lambda f\eqdef (f| \lambda f)_{\cH\otimes \cc^{2}}.
\]
We denote such an operator by $\lambda(\epsilon)$ if all its entries are functions of the selfadjoint operator $\epsilon$.

If  $A= \mat{a}{b}{c}{d}$ we set $A^{\rm diag}= \mat{a}{0}{0}{d}$.  We set also
\beq\label{trouloulou}
\mathcal{T}(t)\defeq 2^{-\12}\mat{\epsilon_{t}^{-\12}}{-\epsilon_{t}^{-\12}}{\epsilon_{t}^{\12}}{\epsilon_{t}^{\12}},\ 
\mathcal{T}^{-1}(t)= 2^{-\12}\mat{\epsilon_{t}^{\12}}{\epsilon_{t}^{-\12}}{-\epsilon_{t}^{\12}}{\epsilon_{t}^{-\12}}.
\eeq
\begin{proposition}\label{p3.1}
 Let $\lambda_{-1}= \lambda_{-1}(\epsilon)$ be a bounded sesquilinear form on $\cA$, $\cB_{-1}$, $\cC_{-1}$  in cases $(A), (B), (C)$ respectively. Then 
 \[
\lambda^{\rm ad}_{1}\defeq \wlim_{T\to +\infty}\hat{U}_{T}(-1, 1)^{*}\lambda_{-1}\hat{U}_{T}(-1, 1) 
\]
exists on $\cA$, $\cB_{1}$, $\cC_{1}$ and
\begin{equation}
\label{e3.18}
\lambda_{1}^{\rm ad}= \mathcal{T}^{-1}(1)^{*}\left(\mathcal{T}(-1)^{*}\lambda_{-1}\mathcal{T}(-1)\right)^{\rm diag} \mathcal{T}^{-1}(1).
\end{equation}
\end{proposition}
\proof 
We first derive an asymptotic expansion in powers of $T^{-1}$ for $\hat{U}_{T}(t,s)= \Texp(\i T\int_{s}^{t}H(\sigma)d\sigma)$ valid for $t, s\in [-1, 1]$. Setting $h= T^{-1}$ this essentially amounts to the construction of WKB solutions of a Schr\"oedinger equation. 

We will find this expansion  by following the construction of a parametrix for the Cauchy problem for Klein-Gordon equations done in \cite{GW14a, GOW16},  taking advantage of the fact that  the equation 
\begin{equation}
\label{e3.14}
(T^{-1}\pe_{t})^{2}\phi+ a(t)\phi=0
\end{equation}
is separable.
We first look for solutions of \eqref{e3.14} of the form $\phi= \Texp(\i T\int_{s}^{t}b_{T}(\sigma)d\sigma)u$ and obtain that  $\phi$ solves \eqref{e3.14} iff $b_{T}(t)$ solves the following Riccati equation:
\begin{equation}
\label{e3.15}
\i T^{-1}\p_{t}b_{T}(t)- b_{T}^{2}(t)+ a(t)=0.
\end{equation}
 We can solve \eqref{e3.15} modulo errors of size $O(T^{-2})$ by
 \beq\label{e3.16c}
b_{T}(t)= \epsilon(t)+ \frac{\i}{2}T^{-1}\p_{t}\ln \epsilon (t).
\eeq
We have then 
\beq\label{e3.16}
\i\p_{t}b_{T}(t)- b_{T}^{2}(t)+ a(t)= T^{-2}(\frac{1}{4} (\p_{t}\ln \epsilon)^{2}- \12\p_{t}^{2}\ln \epsilon)(t).
\eeq
We set  $b_{T}^{+}(t)=b_{T}(t), \ b_{T}^{-}(t)= - b_{T}^{*}(t)$,
\beq\label{e3.16b}
\begin{array}{l}
\mathcal{T}_{T}(t)\defeq  \mat{1}{-1}{b_{T}^{+}}{-b_{T}^{-}}(t)(b_{T}^{+}- b_{T}^{-})^{-\12}(t), \\[2mm]
 \mathcal{T}_{T}^{-1}(t)=  \mat{-b_{T}^{-}}{1}{-b_{T}^{+}}{1}(t)(b_{T}^{+}- b_{T}^{-})^{-\12}(t),
\end{array}
\eeq
and
\[
\hat{U}_{T}(t,s)=: \mathcal{T}_{T}(t)V_{T}(t,s)\mathcal{T}_{T}^{-1}(s).
\]
Mimicking the computations in \cite[Subsect. 6.4]{GOW16},  we easily obtain  that 
\[
V_{T}(t,s)= \Texp(\i T\int_{s}^{t}\hat{H}_{T}(\sigma)d\sigma),
\] for  $\hat{H}_{T}(t)= H^{\rm diag}(t)+ T^{-2}R(t)$, and:
\beq\label{e3.17}
\begin{array}{l}
H^{\rm diag}(t)= \mat{\epsilon(t)}{0}{0}{-\epsilon(t)},\\[2mm]
R_{2}(t)= (2\epsilon)^{-1}(t)(\frac{1}{4} (\p_{t}\ln \epsilon)^{2}- \12\p_{t}^{2}\ln \epsilon)(t)\mat{-1}{1}{-1}{1}.
\end{array}
\eeq
Let us set
\def\diag{{\rm diag}}
\[
V_{T}^{\diag}(t,s)\defeq\Texp(\i T\int_{s}^{t}H^{\rm diag}(\sigma)d\sigma).
\]
In case $A$, $\epsilon_{t}$  is  bounded from below by a strictly positive constant, uniformly for $t\in [-1, 1]$ and we  immediately deduce from  \eqref{e3.17} that  
\[
\|\hat{U}_{T}(t,s)- \mathcal{T}_{T}(t)V^{\diag}_{T}(t,s)\mathcal{T}_{T}^{-1}(s)\|_{B(\cA)}\in O(T^{-1})
\]
uniformly for $t\leq s$, $t,s\in[-1, 1]$. We cannot use  this argument in cases $B, C$ since $0\in \sigma(\epsilon_{t})$, either for $t=-1$ or $t=1$. Instead we use a density argument, that we will explain for case $B$, case $C$ being similar.

From  Lemma \ref{l3.1} we see that the family of sesquilinear forms
\[
\hat{U}_{T}(-1, 1)^{*}\lambda_{-1}\hat{U}_{T}(-1, 1)
\] is bounded on $\cB_{1}$ uniformly for $T\geq 1$. Therefore it suffices to prove \eqref{e3.18} on the dense subspace  $\cS$ defined in \eqref{defdes}.

We have to compute the limit of $(\hat{U}_{T}(-1, 1)f| \lambda_{-1}\hat{U}_{T}(-1, 1)f)_{\cH\otimes \cc^{2}}$ for $f\in \cS$. Since $\hat{U}_{T}(-1, 1)$ and $\lambda_{-1}$ commute with $\epsilon$ we see that if $f= \one_{[\delta, R]}(\epsilon)f$ we can  replace   $\epsilon$ by some function $F(\epsilon)$ such that $\12 \delta\leq F\leq 2R$, $F(\lambda)= \lambda$ on $[\delta, R]$.  Equivalently we can assume that $\epsilon$ is boundedly invertible on $\cH$.

In this way we deduce from Lemma \ref{l3.1} and  \eqref{e3.17} that
\[
\lim_{T\to +\infty}\hat{U}_{T}(-1, 1)f-  \mathcal{T}_{T}(-1)V^{\diag}_{T}(-1,1)\mathcal{T}_{T}^{-1}(1)f=0, \ \forall f\in \cS.
\]
Therefore  we have  as sesquilinear forms on $\cS$:
\[
\begin{array}{rl}
&\hat{U}_{T}(-1, 1)^{*}\lambda_{-1}\hat{U}_{T}(-1, 1)\\[2mm]
=& \mathcal{T}_{T}^{-1}(1)^{*}V_{T}^{\rm diag}(-1, 1)^{*}\hat{\lambda}_{-1, T} V_{T}^{\diag}(-1, 1)\mathcal{T}_{T}^{-1}(1)+ o(T^{0}),
\end{array}
\]
for $\hat{\lambda}_{-1, T}= \mathcal{T}_{T}(-1)^{*}\lambda_{-1}\mathcal{T}_{T}(-1)$.    We have 
\[
V_{T}^{\diag}(t,s)= \mat{u_{T}^{+}(t,s)}{0}{0}{u_{T}^{-}(t,s)}
\]
for
\[
u^{\pm}_{T}(t,s)= \e^{\pm\i  T\int_{s}^{t}\epsilon(\sigma)d\sigma}.
\]
Since $u^{\pm}_{T}(-1, 1)$ is unitary on $\cH$ we can  replace $\hat{\lambda}_{-1, T}$ by 
\beq\label{e3.20}
\hat{\lambda}_{-1}\defeq \mathcal{T}(-1)^{*}\lambda_{-1}\mathcal{T}(-1),
\eeq
where $\mathcal{T}(t)$ is defined in \eqref{trouloulou}.  The error terms will  again be $o(T^{0})$, by \eqref{e3.16c}.
We write  then $\hat{\lambda}_{-1}$ as
\[
\hat{\lambda}_{-1}=\mat{\hat{\lambda}_{-1}^{++}}{\hat{\lambda}_{-1}^{+-}}{\hat{\lambda}_{-1}^{-+}}{\hat{\lambda}_{-1}^{--}}.
\]
Using  that $\hat{\lambda}_{-1}^{\alpha\beta}$ for $\alpha,\beta\in \{+, -\}$ are functions of $\epsilon$, we obtain that:
\[
\begin{array}{rl}
&V^{\diag}_{T}(-1, 1)^{*}\hat{\lambda}_{-1} V_{T}^{\diag}(-1, 1)\\[2mm]
=&\mat{u^{+}_{T}(-1, 1)^{*}\hat{\lambda}_{-1}^{++}u^{+}_{T}(-1, 1)}{u^{+}_{T}(-1, 1)^{*}\hat{\lambda}_{-1}^{+-}u^{-}_{T}(-1, 1)}{u^{-}_{T}(-1, 1)^{*}\hat{\lambda}_{-1}^{-+}u^{+}_{T}(-1, 1)}{u^{-}_{T}(-1, 1)^{*}\hat{\lambda}_{-1}^{--}u^{-}_{T}(-1, 1)}\end{array}
\]
 Now $u^{+}_{T}(t,s)= u^{-}_{T}(t,s)^{*}$ and $\wlim_{T\to +\infty}u^{\pm}_{T}(t,s)=0$ in $\cH$, since the spectrum of $-\Delta_{h}$ is purely absolutely continuous.   This implies that
 \[
\wlim_{T\to +\infty}V^{\diag}_{T}(-1, 1)^{*}\hat{\lambda}_{-1} V_{T}^{\diag}(-1, 1)= \mat{\hat{\lambda}_{-1}^{++}}{0}{0}{\hat{\lambda}_{-1}^{--}}= \hat{\lambda}_{-1}^{\rm diag}
\]
in $\cS$. This completes the proof of the proposition in case $B$, the other cases being similar. \qed
\subsection{Adiabatic limit of vacuum,  thermal  states and Hadamard states}\label{sec4.3}
In the sequel instead of the pair $\lambda^{\pm}$ of  Cauchy surface covariances of some quasi-free state, we will consider only $\lambda^{+}$, (since $\lambda^{-}= \lambda^{+}- q$)  and denote it simply by $\lambda$. The necessary and sufficient condition \eqref{calum} becomes
\begin{equation}
\label{calom}
\lambda\geq0, \ \lambda - q\geq 0.
\end{equation}
 Let $\omega_{-1}$ be a quasi-free state for the Klein-Gordon operator at time $t=-1$, ie $P_{-1}=\pe_{t}^{2}- \Delta_{h}+ m^{2}_{-1}$ and let $\lambda_{-1}$ its covariance at time $t=-1$. 
In order to be able to apply Prop. \ref{p3.1} to study the adiabatic limit $\lambda_{1}^{\rm ad}$ of $\lambda_{-1}$ we need that the following properties are satisfied:
\ben
\item $\lambda_{-1}$ is bounded on $\cA$, resp. $\cB_{-1}$, $\cC_{-1}$;
\item $\coinf(\Sigma)\otimes \cc^{2}\subset\cA$, resp. $\cB_{-1}$,  $\cC_{-1}$ continuously;
\item $\coinf(\Sigma)\otimes \cc^{2}\subset\cA$, resp. $\cB_{1}$,  $\cC_{1}$ continuously.
\een
In fact (1) is needed to obtain the existence of the adiabatic limit $\lambda^{\rm ad}_{1}$ on $\cA$, resp. $\cB_{1}$, $\cC_{1}$, while (2) and (3) imply that the initial covariance $\lambda_{-1}$ and final covariance $\lambda_{1}^{\rm ad}$ are well defined on $\coinf(\Sigma)\otimes \cc^{2}$.

In particular since \eqref{calom} is automatically satisfied by $\lambda_{1}^{\rm ad}$, $\lambda^{\rm ad}_{1}$ is   the covariance  at time $t=1$ of a quasi-free state $\omega_{1}^{\rm ad}$ for the Klein-Gordon operator at time $t=1$, ie $P_{1}= \pe_{t}^{2}- \Delta_{h}+ m^{2}_{1}$.

\subsubsection{Adiabatic limit of thermal states (case $A$)}
We assume we are in  case $A$ and take as initial state the $\beta-${\em KMS state} at time $t=-1$, given by the covariance:
\[
\lambda_{-1}^{\beta} = \12\mat{\epsilon_{-1}\coth (\beta\epsilon_{-1}/2)}{\one}{\one}{\epsilon^{-1}_{-1}\coth(\beta\epsilon_{-1}/2)}.
\]
\begin{theorem}\label{th4.1}
The adiabatic limit
 \[
\lambda_{1}^{\beta, {\rm ad}}= \wlim_{T\to +\infty}\hat{U}_{T}(-1, 1)^{*}\lambda_{-1}^{\beta}\hat{U}_{T}(-1, 1)
\]
exists on $\coinf(\Sigma)\otimes \cc^{2}$.  The adiabatic limit state $\omega_{1}^{\beta, {\rm ad}}$ is {\em not} the $\beta-$KMS state at time $t=1$, unless $m_{1}= m_{-1}$.
\end{theorem}
\proof 
Properties (1), (2), (3)  are immediate for the space $\cA$, using that the mass of the field is strictly positive.   A routine computation  shows that the limit covariance $\lambda_{1}^{\beta,{\rm  ad}}$ in Prop. \ref{p3.1} equals:
\beq\label{poldu}
\lambda_{1}^{\beta,{\rm ad}}=\12\mat{\epsilon_{1}\coth (\beta\epsilon_{-1}/2)}{\one}{\one}{\epsilon^{-1}_{1}\coth(\beta\epsilon_{-1}/2)}.
\eeq
This is not the covariance of the $\beta-$KMS state at time $t=1$, unless $m_{1}= m_{-1}$. \qed
\begin{remark}
  The instability of KMS states under adiabatic limits can be related to the  failure of the  return to equilibrium property analyzed in \cite{DrFaPi16}. In this paper the authors 
 consider a couple of KMS states $\omega^\beta$, $\omega^\beta_V$ with respect to different dynamics $\tau$,$\tau^V$.
Here, $\tau^V$ is the one-parameter group of $*$-automorphism obtained by perturbing the dynamics $\tau$ with a self-adjoint element $V$.
The state $\omega^\beta$ is said to satisfy the return to equilibrium property if $\textrm{w-}\lim_{t\to\infty}\omega^\beta\circ\tau^V_t=\omega^\beta_V$.
In \cite{DrFaPi16} it has been shown that, for quantum fields, such a property is linked to the support properties of $V$.
Actually, if the spatial support of $V$ is compact, then $\omega^\beta$ satisfies the return to equilibrium property, while if $V$ has non-compact spatial support this is not the case.\\
In our case  the adiabatic limit $\wlim_{T\to +\infty}\hat{U}_{T}(-1, 1)^{*}\lambda_{-1}^{\beta}\hat{U}_{T}(-1, 1)$ can be related with $\lim_{t\to\infty}\omega^\beta\circ\tau^V_t$, by identifying the perturbation $V$ with the quadratic perturbation $\int m^2\chi(t)\phi^2(x) dVol_g$, which is not of compact spatial support. 
\end{remark}
\subsubsection{The infrared problem}
 To verify properties (2) (3), in particular the inclusions $\coinf(\Sigma)\otimes \cc^{2}\subset \cB_{-1}, \cC_{1}$, one is faced with a version of the {\em infrared problem}, ie the fact that $0\in \sigma(\epsilon)$. In the lemma below we give a sufficient condition for (2), (3) which is easy to verify in applications. 
\begin{lemma}\label{adiabato}
Assume that:
 \[
(IR)\ \hbox{ there exists a  continuous function }c: \Sigma\to \rr, \ c(x)>0\hbox{ such that }-\Delta_{h} \geq c^{-2}(x).
\]
Then (2), (3) are satisfied.
\end{lemma}
\begin{remark}
 If $\Sigma= \rr^{d}$ and the metric $h$ satisfies
\beq\label{e3.16d}
h_{ij}(x)\geq C \delta_{ij}, \ \p^{\alpha}_{x}h_{ij}(x)\hbox{ bounded  for all }\alpha\in \nn^{d},
\eeq
then $(IR)$ holds for $c(x)= C \langle x\rangle$, see \cite[Prop. A2]{GGH}.
\end{remark}
\proof 
We immediately see that  if
\beq\label{e3.201}
\coinf(\Sigma)\subset \Dom \epsilon^{-\12}\cap \Dom \langle \epsilon\rangle^{\12}\epsilon^{-1}\cap \Dom \langle \epsilon\rangle \epsilon^{-\12}
\eeq
then properties (2)  and (3) are satisfied. From functional calculus \eqref{e3.201} holds if $\coinf(\Sigma)\subset \Dom \epsilon^{-1}$.  
Setting $A= c^{-2}(x)$, $B= -\Delta_{h}= \epsilon^{2}$ we have $0<A\leq B$, which by definition means that $\Ker A= \{0\}$, $\Dom B^{\12}\subset \Dom A^{\12}$ and $(u|Au)\leq (u|Bu)$ for $u\in \Dom B^{\12}$.
By \cite[Thm. V.2.21]{K} this implies that $0<(B+\delta)^{-1}\leq (A+\delta)^{-1}$ for any $\delta>0$. Letting $\delta\to 0^{+}$ we obtain $0<B^{-1}\leq A^{-1}$ ie $\Dom c\subset \Dom \epsilon^{-1}$, 
which completes the proof since $\coinf(\Sigma)\subset \Dom c$ \qed

\subsubsection{Adiabatic limit of vacuum state (cases $B$, $C$)}
We assume that we are in case $B$ or $C$ and take as initial state the  {\em vacuum state} at time $t=-1$  given by the covariance
\[
\lambda_{-1}^{\rm vac}=\12\mat{\epsilon_{-1}}{\one}{\one}{\epsilon^{-1}_{-1}}.
\] 
\begin{theorem}\label{th4.2}
Assume that $(IR)$ holds.  Then 
the adiabatic limit
 \[
\lambda_{1}^{{\rm vac}, {\rm ad}}= \wlim_{T\to +\infty}\hat{U}_{T}(-1, 1)^{*}\lambda_{-1}^{\rm vac}\hat{U}_{T}(-1, 1)
\]
exists on $\coinf(\Sigma)\otimes \cc^{2}$.  The adiabatic limit state $\omega_{1}^{{\rm vac}, {\rm ad}}$ is   the vacuum  state at time $t=1$.
\end{theorem}
 \proof Property (1) holds by direct computation and  (2), (3) hold by  Lemma \ref{adiabato}.  We apply then Prop. \ref{p3.1}.  The same computation as Thm. \ref{th4.1}, which amounts to set $\beta= +\infty$ in \eqref{poldu}, shows that $\lambda_{1}^{{\rm vac}, {\rm ad}}$ is the covariance of the vacuum state at time $t=1$. \qed
 \subsubsection{Adiabatic limit for a class of Hadamard states (cases A, B, C)}
 
 We now take as initial state a {\em Hadamard state} at  time $t=-1$, whose covariance $\lambda_{-1}$ is a function of $\epsilon$, as in Prop. \ref{p3.1}.  This corresponds exactly to a Hadamard state obtained by mode decomposition arguments. 
 
Let us first discuss the  form of the covariance $\lambda_{-1}$.  

Recall that we have set $\mathcal{T}(-1)^{*}\lambda_{-1}\mathcal{T}(-1)\eqdef \hat{\lambda}_{-1}$. Using that 
\[
\mathcal{T}(-1)^{*}q\mathcal{T}(-1)= \mat{\one}{0}{0}{-1}\eqdef \hat{q},
\]
  the positivity condition \eqref{ef.1} becomes
\[
\hat{\lambda}_{-1}\geq 0, \hat{\lambda}_{-1}\geq \hat{q}.
\]
This is satisfied if 
\[
\hat{\lambda}_{-1}= \mat{\one + b^{*}b}{b^{*}dc}{c^{*}db}{c^{*}c},
\]
for $b, c, d\in L(\cH)$ and $\| d\|_{B(\cH)}\leq 1$, see eg \cite[Prop. 7.4]{GW14a}.   The operators $b, c, d$ should be functions of $\epsilon$, ie $b= b(\epsilon), c= c(\epsilon), d= d(\epsilon)$ for Borel measurable functions $b,c,d: \rr^{+}\to \rr$, the requirement $\| d(\epsilon)\|\leq 1$ being insured if $|d(s)|\leq 1$ for $s\in \rr^{+}$.

Finally  $\lambda_{-1}$ should be a Hadamard state, which is ensured if $\lambda_{-1}- \lambda_{-1}^{\rm vac}$ is infinitely smoothing.  Using the ellipticity of $- \Delta_{h}$, this is the case if
\[
b(s), c(s)\in O(\langle s\rangle^{-\infty}).
\]

We now discuss the conditions (1), (2), (3) in the beginning of Subsect. \ref{sec4.3}. We saw in Lemma \ref{adiabato} that (2), (3) are satisfied if condition $(IR)$ holds, so it remains to discuss condition (1), ie the fact that $\lambda_{-1}$ is bounded on $\cA$, $\cB_{-1}$ or $\cC_{-1}$.  Equivalently  if $\hat{\cA}, \hat{\cB}_{-1}, \hat{\cC}_{-1}$  are the images of $\cA, \cB_{-1},\cC_{-1}$ under $\mathcal{T}(-1)^{-1}$, $\hat{\lambda}_{-1}$ should be bounded on   $\hat{\cA}, \hat{\cB}_{-1}, \hat{\cC}_{-1}$, in cases  $(A), (B), (C)$.

  An easy computation yields that:
\[
\hat{\cA}= \hat{\mathcal{B}}_{-1}= \hat{\mathcal{C}}_{-1}=\mathcal{H}\oplus \mathcal{H},
\]
hence condition (1) is satisfied if $b,c,d$ are bounded functions. Summarizing we impose the following condition on the initial covariance:
\begin{equation}
\label{taratata}
\hat{\lambda}_{-1}= \mat{\one + b^{*}b(\epsilon)}{b^{*}dc(\epsilon)}{c^{*}db(\epsilon)}{c^{*}c(\epsilon)}, \hbox{for }b, c, d: \rr^{+}\to \rr,  b(s), c(s)\in O(\langle s\rangle^{-\infty}), |d(s)|\leq 1.
\end{equation}

  \begin{theorem}\label{th4.3} Let $\omega_{-1}$ be a Hadamard state at time $t= -1$, whose covariance $\lambda_{-1}$ is such that $\hat{\lambda}_{-1}$ satisfies \eqref{taratata}. In cases (B), (C) we assume moreover  condition (IR).
Then the adiabatic limit
  \[
\lambda_{1}^{ {\rm ad}}= \wlim_{T\to +\infty}\hat{U}_{T}(-1, 1)^{*}\lambda_{-1}^{\rm Had}\hat{U}_{T}(-1, 1)
\]
exists on $\coinf(\Sigma)\otimes \cc^{2}$.  The adiabatic limit state  $\omega_{1}^{\rm ad}$ is a Hadamard state at time $t=1$.
\end{theorem}
\proof   the existence of $\lambda_{1}^{\rm ad}$ follows from Prop. \ref{p3.1}.  We obtain that 
\[
(\mathcal{T}(-1)^{*}\lambda_{-1}\mathcal{T}(-1))^{\rm diag}= \hat{\lambda}_{-1}^{\rm diag}= \mat{1+ |b|^{2}(\epsilon)}{0}{0}{|c|^{2}(\epsilon)}.
\]
It follows that $\lambda_{1}^{\rm ad}= \lambda_{1}^{\rm vac}+ r$, where 
\[
r= \mat{\epsilon_{1}(|b|^{2}(\epsilon)+ |c|^{2}(\epsilon)}{|b|^{2}(\epsilon) - |c|^{2}(\epsilon)}{|b|^{2}(\epsilon)- |c|^{2}(\epsilon)}{\epsilon_{1}^{-1}(|b|^{2}(\epsilon)+ |c|^{2}(\epsilon)}.
\]
Using that  by $(IR)$ $\coinf(\Sigma)\subset \Dom \epsilon_{1}^{-1}$ and  the fact that $b(s), c(s)\in O(\langle s\rangle^{-\infty})$ we obtain that $r$ is smoothing, hence 
 $\omega_{1}^{\rm ad}$ is Hadamard. \qed

\end{document}